\documentclass[lettersize,journal]{IEEEtran}
\usepackage{amsmath,amsfonts}
\usepackage{algorithmic}
\usepackage{algorithm}
\usepackage{array}
\usepackage[caption=false,font=normalsize,labelfont=sf,textfont=sf]{subfig}
\usepackage{textcomp}
\usepackage{stfloats}
\usepackage{url}
\usepackage{verbatim}
\usepackage{soul}
\usepackage{graphicx}
\usepackage{cite}
\usepackage{multirow}
\usepackage{epsfig}
\usepackage{color}

\renewenvironment{thebibliography}[1]{%
   \begin{oldthebibliography}{#1}%
     \setlength{\itemsep}{-.3ex}%
}%
{%
   \end{oldthebibliography}%
}

\usepackage[belowskip=-12pt,aboveskip=0pt]{caption}
\usepackage[utf8]{inputenc}
\usepackage{array}
    \newcolumntype{P}[1]{>{\centering\arraybackslash}p{#1}}
    \newcolumntype{M}[1]{>{\centering\arraybackslash}m{#1}}

\usepackage{stackengine}

\usepackage{booktabs}
\newcommand{\ra}[1]{\renewcommand{\arraystretch}{#1}}

\usepackage{caption} 
\def\BibTeX{{\rm B\kern-.05em{\sc i\kern-.025em b}\kern-.08em
    T\kern-.1667em\lower.7ex\hbox{E}\kern-.125emX}}
\usepackage{balance}

\begin{document}

\title{Understanding Security in Smart City Domains From the ANT-centric Perspective}

\author{Jiani Fan, Wenzhuo Yang, Ziyao Liu, Jiawen Kang, Dusit Niyato,~\IEEEmembership{Fellow,~IEEE},  Kwok-Yan Lam,~\IEEEmembership{Senior Member,~IEEE}, and Hongyang Du
\thanks{Manuscript received February 05, 2022; This research is supported by the Singapore National Research Foundation, under its Strategic Capability Research Centres Funding Initiative. Any opinions, findings, and conclusions or recommendations expressed in this material are those of the author(s) and do not necessarily reflect the views of National Research Foundation, Singapore; Jiani Fan's research is partly supported by Alibaba Group through Alibaba Innovative Research (AIR) Program and Alibaba-NTU Singapore Joint Research Institute (JRI), Nanyang Technological University, Singapore. \emph{(Corresponding author: wenzhuo001@e.ntu.edu.sg, Wenzhuo Yang.)}
}
}

\markboth{IEEE INTERNET OF THINGS JOURNAL, ~VOL.~XX, NO.~XX, FEBRUARY, ~2022}%
{}

\maketitle

\begin{abstract}
A city is a large human settlement that serves the people who live there, and a smart city is a concept of how cities might better serve their residents through new forms of technology. 
In this paper, we focus on four major smart city domains according to Maslow's hierarchy of needs: smart utility, smart transportation, smart homes, and smart healthcare. Numerous IoT applications have been developed to achieve the intelligence that we desire in our smart domains, ranging from personal gadgets such as health trackers and smart watches to large-scale industrial IoT systems such as nuclear and energy management systems. However, many of the existing smart city IoT solutions can be made better by considering the suitability of their security strategies. Inappropriate system security designs generally occur in two scenarios: first, system designers recognize the importance of security but are unsure of where, when, or how to implement it; and second, system designers try to fit traditional security designs to meet the smart city security context. Thus, the objective of this paper is to provide application designers with the missing security link they may need in order to improve their security designs. By evaluating the specific context of each smart city domain and the context-specific security requirements, we aim to provide directions on when, where, and how they should implement security strategies and the possible security challenges they need to consider. In addition, we present a new perspective on security issues in smart cities from a data-centric viewpoint by referring to the reference architecture, the Activity-Network-Things (ANT)-centric architecture. This architecture is built upon the concept of ``security in a zero-trust environment," to achieve end-to-end data security. By doing so, we reduce the security risks posed by new system interactions or unanticipated user behaviors while avoiding the hassle of regularly upgrading security models.

\end{abstract}


\section{Introduction}

    \IEEEPARstart{S}mart city is a new paradigm that utilizes various advanced technologies, especially artificial intelligence, to improve urban conditions and make urban life more convenient. At this moment, there is no clear concept of what a smart city should look like, but we can imagine its potential in every possible way. After all, every smart city is the result of the joint choice and vision of all parties involved. 
    
    According to Maslow's hierarchy of needs \cite{mcleod2007maslow}, the basic needs of human includes food, water, warmth, rest, safety and security, which form the basic layout of smart cities. In any smart city, smart utility systems are required for the efficient provision of power and to enable the operation of other smart systems; Smart transportation systems are in charge of distributing resources quickly across cities; Smart homes are intended to provide warm, comfortable shelters for city dwellers; Smart healthcare systems are responsible for providing adequate health security. Thus, we focus on four major smart city domains in this paper: smart utility, smart transportation, smart homes, and smart healthcare. Nevertheless, all smart city domains will be empowered through the deployment of extensive Internet of Things (IoT) device networks.
    
    IoT refers to the network of interconnected devices that enables cyberspace to interact with the environment and infrastructure\cite{IOTdefine}. The wireless characteristics of data communication in most IoT applications, as well as its ease of implementation, have accelerated the adoption of IoT systems in recent years. According to Statista \cite{statista}, the number of IoT devices is expected to reach 30.9 billion by 2025, triple the number of non-IoT devices by then. With the support of IoT technologies and wireless communication protocols, our cities are capable of building extensive data networks that incorporate a large number of heterogeneous devices that participate in the data collection and decision-making process. This massive data backbone formed by IoT devices allows systems to perform real-time environmental monitoring and actuating. With such capabilities, our cities will be able to generate service intelligence and provide a higher standard of living without jeopardizing sustainability. 
    
    Numerous IoT applications have been developed to achieve the intelligence required for our smart cities, ranging from personal gadgets such as health trackers and smart watches to large-scale industrial IoT systems such as public transportation and energy management systems. However, many of the existing smart city IoT solutions must be carefully designed by considering the security strategies. Inappropriate system security designs generally occur in two scenarios: First, the system designers recognize the importance of security but are unsure of where, when, or how to implement it. Second, the system designers try to fit traditional security designs to meet the smart city security context without taking specific characteristics of the smart city applications and IoT systems into account. The former necessitates a close examination of the specific security requirements on a context-by-context basis, which will be elaborated further in Section \ref{sec: context}. Moreover, the latter requires a thorough understanding of the characteristics of smart city applications and their security challenges.
    
    For example, a common characteristic of smart city applications is that they are data-driven and rely on constant ubiquitous sensing of the environment. This creates a significant concern regarding compatibility issues with big data due to the high volume, velocity, and variety of data flow~\cite{bigDataChallenge21,bigdataheterogeneity17}. Maintaining data integrity and confidentiality on such a massive scale needs a large amount of computation and storage resources, and failing to do so will have disastrous consequences for the usability of smart systems. Thus, how to best implement adequate data security protection for various types and stages of data is a difficult challenge that smart city applications must address.
    
    Another common characteristic is user-centric design in smart city applications. Besides common objectives, such as optimizing resource utilization and increasing system efficiency in conventional systems, improving user experience will be a main objective for standard smart city applications. A large amount of sensitive private data is collected to understand user preferences and behavior to provide personalized service and designs\cite{smarthomeauthentication20}. This has raised significant privacy concerns about the use of private data. Any smart city application dealing with such data will require substantial efforts to protect and regulate them.
    
    Furthermore, a common smart city application is expected to deal with devices of varied capabilities and resource constraints and frequent data interchange inside the system and with third parties, imposing high interoperability requirements on the smart city application. This has created a new security context for smart city applications. By a security context, we mean the scope of responsibility that the security mechanism must protect. For example, conventional system security solely protects IT components such as the software system and servers. However, IoT security in a smart city may also have to safeguard operational components such as sensors and actuators that are physically installed\cite{fromittoot16}. Common security challenges include integrity, confidentiality, availability, decentralization, authentication, secure communication, end-to-end security, identity management, access control, privacy, system resilience, scalability, and interoperability \cite{asurveyoniotsecurityrequirements, securityRequirementsIoT20}. With a changing security context, we will face new security challenges that we may not have anticipated. For example, we have to consider the different wireless communication protocols adopted by IoT devices, where incompatibility between different protocols can lead to serious communication security flaws; and the deployment environment, in which IoT devices may operate in uncontrolled environments such as public and open spaces, as opposed to the controlled environments in which they have previously operated. Thus, establishing security mechanisms based on a ``zero-trust" environment is essential for securing heterogeneous interactions. As a result, conventional system security designs may not be appropriate to be directly applied to smart city applications.

    Hence, the objective of this paper is to provide application designers with the missing security link they may need in order to improve their security designs. By evaluating the specific context of each smart city domain and the context-specific security requirements, we aim to provide directions on when, where, and how they should implement security strategies and the possible security challenges they need to consider.

    In this paper, we look at the cybersecurity requirements and challenges for the smart cities from a data-centric perspective by referring to a security reference architecture, i.e., the Activity-Network-Things (ANT)-centric architecture \cite{ANTcentric21}. The ``Activity-centric" view establishes the system context and highlights critical nodes in the system where sensitive data is stored and processed. The ``Network-centric" view investigates the connectivity between heterogeneous system components and the sensitive data flow. The ``Things-centric" view aids in determining whether the capabilities of end-devices or physical things in terms of communication, computation, and energy are adequate for the execution of adequate cryptographic operations. 
    
    There are many benefits of adopting this approach.
    Firstly, this architecture is built on the notion of ``security in a zero-trust environment", identifying and assessing all attack surfaces under the complex heterogeneous smart city IoT systems. As a result, there is less reliance on the involvement of trusted agents, reducing the risk of insider attacks and taking into account data protection in the broader ecosystem.
    Furthermore, unlike traditional architecture, ANT-centric architecture is centered on the concept of end-to-end data security, in which we attempt to safeguard sensitive data flows and their interactions rather than bounded system components. By doing so, we avoid the trouble of constantly upgrading the security models as new system interactions and functions are implemented.
    In addition, with this security reference architecture, we can identify critical components of the system that handle sensitive data and understand the data interactions with system components. This provides a means for us to determine the sensitivity of data as well as the criticality of activities evolving around this data.
    Lastly, we chose the ANT architecture also because of its flexibility to suit any IoT system and its comprehensive representation of the IoT systems from the architectural perspectives of device, internet and semantic. Thus, security requirements developed using this security reference architecture can be easily applied across smart city applications.

    The main contributions of this paper are summarized as follows:
    \begin{itemize}
    \item \textcolor{black}{We present a new perspective on security issues in smart cities from a data-centric viewpoint by referring to the ANT reference architecture. This aims to provide a different view towards approaching privacy and security issues in a smart city context.}
    \item \textcolor{black}{We investigate the concept of a smart smart city holistically, beginning with end applications to abstract shared properties and highlighting potential privacy and security challenges. By doing so, we hope to broaden readers' understanding of the background and characteristics of smart cities.}
    \item \textcolor{black}{We evaluate the specific context of each smart city domain and the context-specific security requirements from the ANT-centric perspective. By doing so, we aim to provide application designers with the missing security link they may need in order to improve their security designs and directions on when, where, and how they should implement security strategies and the possible security challenges they need to consider.}
    \item \textcolor{black}{From the ANT-centric standpoint, we present the set of challenges that current security solutions for smart city applications will encounter in meeting this set of requirements and provide a new research perspective towards the evaluation of security frameworks and design. These could also serve as future research directions where research can be aimed at fulfilling these challenges in existing frameworks or giving insights into future research directions.}
    \item \textcolor{black}{We discuss the potential prospects for IoT and IoT security in smart cities, such as cognitive IoT and the adoption of machine learning and artificial intelligence.}
    \end{itemize}


 \begin{table}[h]
    \centering 
    \begin{tabular}{ | p{0.6cm} | p{1.15cm} | p{5.3cm}| } 
      \hline
      \hline
      \textbf{Year} & \textbf{Reference} & \textcolor{black}{\textbf{Contribution}}  \\ 
      \hline
      \hline
      2019 & \cite{currentresearchiot} & \textcolor{black}{Presented IoT attack vectors and summarize research according to the security technique used.}\\ 
      \hline
      2021, 2019, 2020, 2022&\cite{asurveyoniotsecurityrequirements,neshenko2019demystifying,securityconsiderationsforiot},\cite{anoverviewofsecurity} & \textcolor{black}{Surveyed IoT security threats, attacks, and vulnerabilities and risk mitigation methods.}\\ 
      \hline
      2019 &\cite{theeffectofiot} & \textcolor{black}{Presented ``IoT features" for representing IoT security challenges, relating each feature into IoT threat, challenge and opportunity.} \\ 
      \hline
      2019 &\cite{asurveyoniotsecurity} & \textcolor{black}{Surveyed IoT security from a data perspective, considering data life cycle.} \\ 
      \hline
      2018 &\cite{securityandprivacyinsmart} & \textcolor{black}{Surveyed IoT security in smart city from the viewpoint of related disciplines such as cryptography, blockchain, biometrics, machine learning.} \\ 
      \hline
      2019 &\cite{securityandprivacyofs} & \textcolor{black}{Presented different types of attacks and techniques, broad classification of security challenges into IoT-based and cloud-based under smart city landscape.} \\ 
      \hline
      2021 &\cite{haque2021conceptualizing} & \textcolor{black}{Provided holistic overview and conceptual development of smart city.}\\ 
      \hline
      - & This work & \textcolor{black}{We present a new approach to smart city security based on the ANT-centric security reference architecture, evaluate the specific context of each smart city domain and the context-specific security requirements, and provide context-specific guidance on when, where, and how system designs can implement security strategies. Furthermore, we examine the challenges that current security proposals face in meeting these requirements, as well as our vision for future IoT and IoT security.} \\ 
      \hline
    \end{tabular}
    \caption{\textcolor{black}{Comparing our paper to other related works.}}
    \label{tab: tech1111}
    \end{table}

\begin{figure*}[htb]
    \begin{center}
        \includegraphics[width=0.8\linewidth]{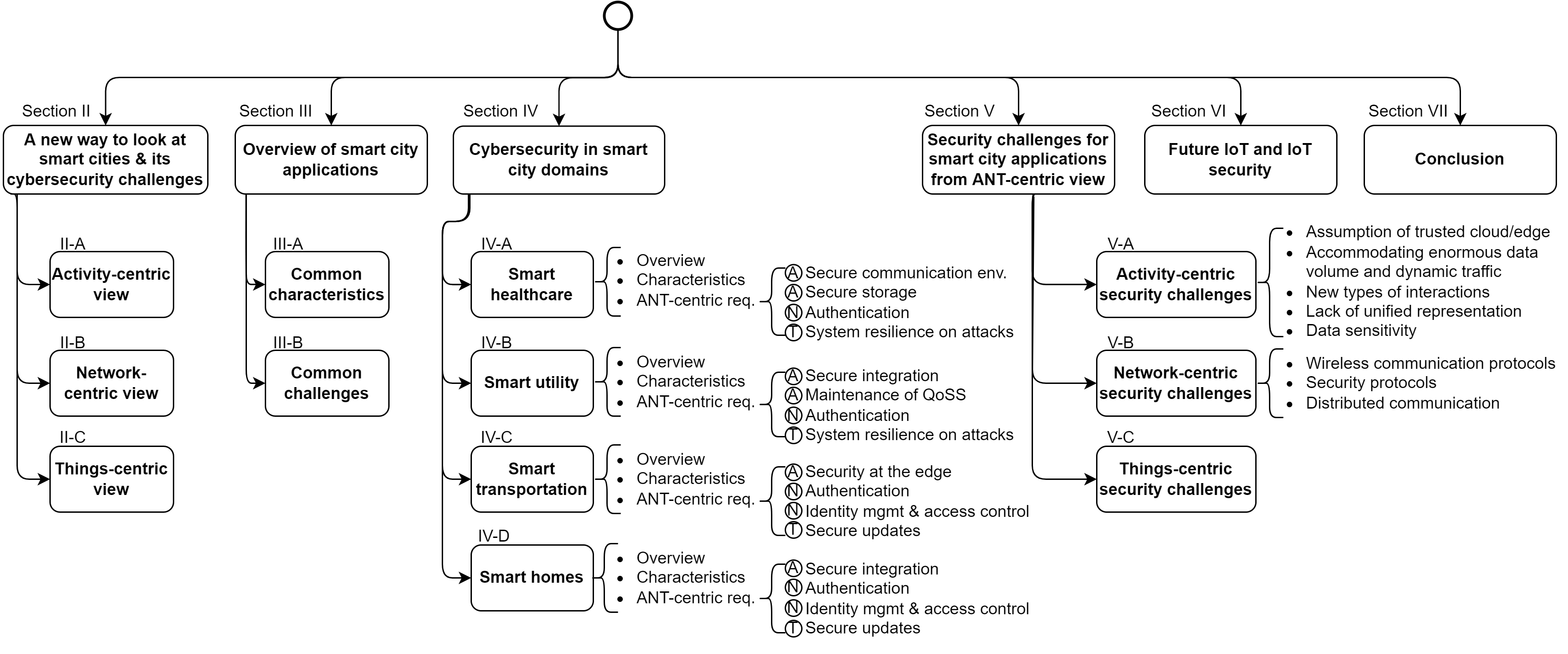}
        \caption{Organization of this paper.}
        \label{fig:flowchart}
    \end{center}
    \end{figure*}

    \textcolor{black}{The structure of this paper is in accordance with the following: In Section \ref{sec: ant}, we explain the ANT reference architecture and how to view security issues in smart cities from a data-centric perspective by referring to the architecture.}
    \textcolor{black}{In Section \ref{sec: overview}, we study the concept of a smart smart city holistically and highlight some of the privacy and security challenges that smart city applications will face.}
    \textcolor{black}{In Section \ref{sec: context}, we evaluate the specific context of each smart city domain and their security requirements from an ANT-centric perspective, presenting existing security proposals and some of their shortcomings.}
    \textcolor{black}{In Section \ref{sec: challenges}, we investigate the challenges and gaps that current security solutions have to overcome to meet the set of security requirements from an ANT-centric perspective.}
    \textcolor{black}{In Section \ref{sec: future}, we discuss the potential prospects for IoT and IoT security in smart cities, such as cognitive IoT and the adoption of machine learning and artificial intelligence.}
    \textcolor{black}{In Section \ref{sec: conclusion}, we conclude our work in this paper.} The organization of this paper is also presented via Figure \ref{fig:flowchart}.
    
\section{ANT-centric: A New Way To Look At Smart Cities and Its Cybersecurity Challenges} \label{sec: ant}
In this section, we aim to give a new perspective on smart cities and their cybersecurity challenges to facilitate a deeper understanding of the security requirements needed for designing future smart city solutions. Most of the existing security survey papers for IoT-enabled smart cities consider the security and privacy challenges based on the device layer, network layer, software layer to application layer \cite{securityandprivacyinsmart}. In this survey, we look at the cybersecurity requirements and challenges for the smart cities from a data-centric perspective by referring to a security reference architecture, the Activity-Network-Things (ANT)-centric architecture \cite{ANTcentric21}. This architecture is built on the notion of ``security in a zero-trust environment", identifying and assessing all attack surfaces. Therefore, by referencing to this security reference architecture, security requirements in smart cities can be examined and designed more thoroughly. We choose the ANT architecture also because of its flexibility to suit any IoT system. Furthermore, approaching the security requirements and problems for smart cities from a data-centric standpoint makes it more natural and easier to ensure the privacy of sensitive data throughout its life cycle. From here, we will explain the ANT reference architecture and how to view security issues in smart cities from a data-centric perspective by referring to the architecture.

\subsection{Activity-Centric View}
An important concept in the ANT architecture is ``critical activities", which affects the customization of security control countermeasures in the IoT system. Critical activities include but are not limited to Sensing and Actuating (SA), Preprocessing (PP), Processing (DA), Crypto Endpoint (CE), Network transport (NT), Storage (ST), Command and Control (CC), and Interfacing with external systems (IO) \cite{ANTcentric21}. More critical activities can be customized based on the features of different IoT systems. 
The activity-centric view is a new way to look at the generation, collection, transmission, processing, and aggregation of sensitive data in smart cities. The security requirements can be considered during the whole life cycle of the data flow. End-to-end secure data flow and storage are required to ensure data quality and availability in the IoT-enabled smart cities.

\subsection{Network-Centric View} 
There are many types of networks and protocols in different smart city applications, which makes the network-centric view also important to think about when planning the security requirements. The communication architecture in smart cities can be categorized as public, private, or protected to facilitate risk assessment. The network-centric view makes it easier to identify and assess security policies in the network for critical nodes that are related to critical activities. In the smart environment, there is frequent sensing and wireless communication among different components of various smart city applications, even in the open public space. Therefore, end-to-end secure communication on the network through identity authentication, identity management, and the construction of a secure communication environment is necessary to guarantee the security and privacy of sensitive data.
\subsection{Things-Centric View}
IoT-enabled smart cities are fabrics of various devices, systems, and other physical things.
The things-centric view in ANT architecture inspire us to consider security problems closely related to the devices itself. New methods for secure software updates to accommodate the features of future smart cities are needed. In addition, security physical things in the IoT systems should have the capability to take proactive detection for intrusion in the early stage of the cyber attacks and should also own the ability to heal quickly after being attacked.
Therefore, robust and resilient physical things are essential for smart city applications to secure the generation, collection, transmission, processing, and aggregation processes of sensitive data.
In the following sections, we will present an overview of the smart city paradigm and a summary of the shared characteristics among smart city applications and highlight some of the common security and privacy challenges.

\section{Overview of Smart City Applications \label{sec: overview}}

    \begin{figure*}[htb]
    \begin{center}
        \includegraphics[width=0.8\linewidth]{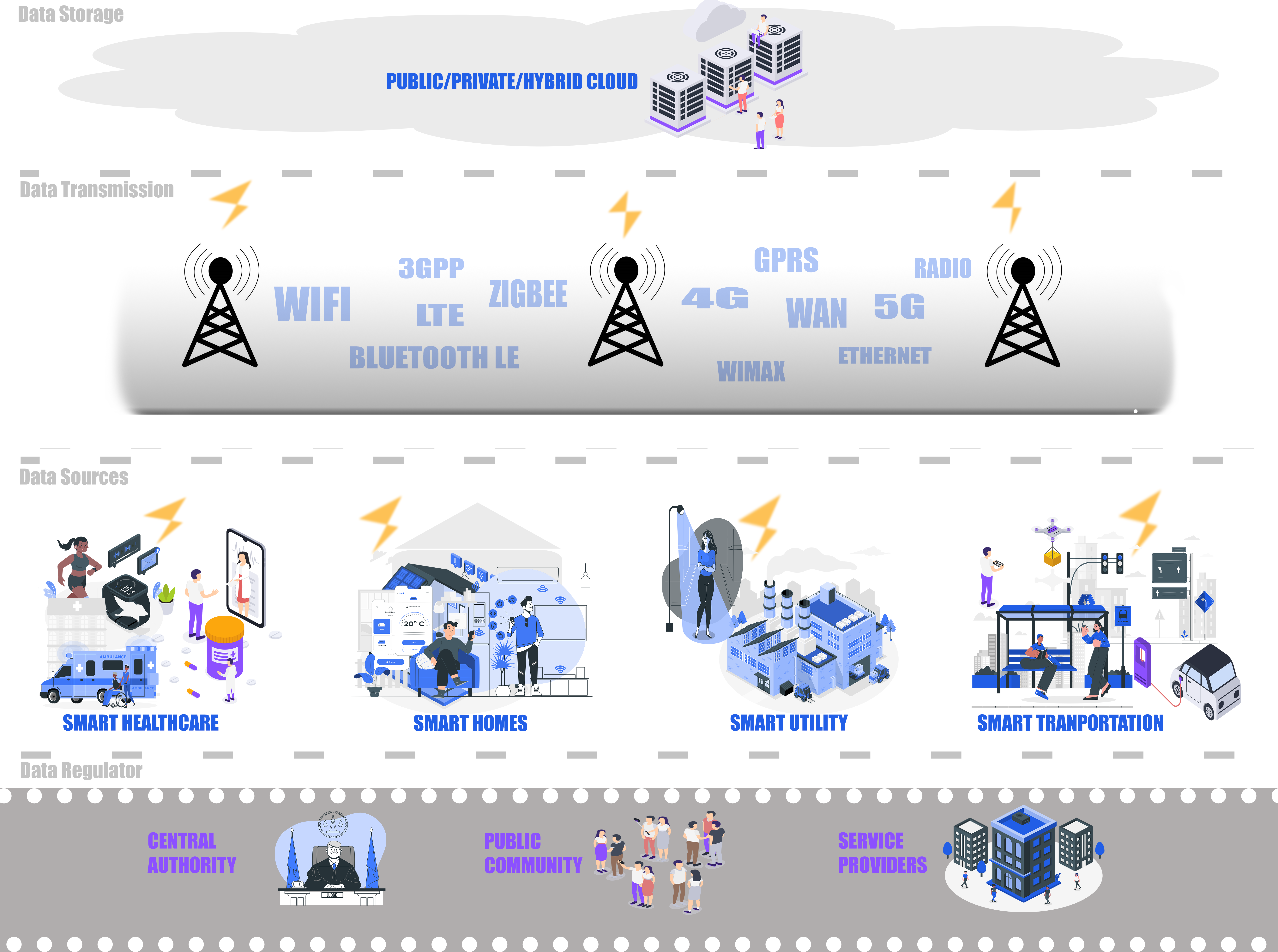}
        \caption{An illustration of major smart city developments.}
        \label{fig:smart city overview}
    \end{center}
    \end{figure*}
    
    \textcolor{black}{In this section, we first present an overview of the smart city paradigm and then introduce the major domains of smart cities. With the smart city landscape in mind, we will present a summary of the shared characteristics among smart city applications and highlight some of the common security and privacy challenges.}
    
    To begin, to provide a comprehensive presentation of the smart city landscape, we show an illustration of the major smart city domains in Fig.~\ref{fig:smart city overview}\footnote{This diagram contains illustrations from https://storyset.com}. In this scenario, central authority, public community, and service providers are the major players involved in controlling and maintaining the city. \textcolor{black}{Variety of smart devices and applications consists of the key elements of these major players.} Smart devices and applications create a bidirectional data network in which any participant can ingest or contribute data. These data exchanges can be done by various wireless communication protocols, and the cloud is likely to be used to handle such massive data traffic.
   \textcolor{black}{Under this broad image of the smart city landscape, we consider the most important smart city domains: smart healthcare, smart utility, smart transportation and smart homes\cite{haque2021conceptualizing}. In Table \ref{tab: examples}, we provide a summary table of examples of smart city applications in these sectors as well as the key security or privacy issues they investigate.}

   \begin{table*}[hbt!]
    \centering
    
    \scalebox{0.92}{
    
    \ra{1.3}
    \begin{tabular}{@{}rrrrrrrr@{}}\toprule
    
    & \multicolumn{7}{c}{Major Issues Studied} \\
    \cmidrule{2-8}
    & Confidentiality & Integrity & Availability & Privacy & Interoperability & Efficiency & System Monitoring\\ \midrule
    \textbf{Smart Healthcare}\\
    TMIS & \cite{healthmonitor21, Awasthi,Tan3factor,Mishra14,Yan13,Amin15,zhang17,CBMSHealth21} & \cite{healthmonitor21, Awasthi,Tan3factor,Mishra14,Yan13,Amin15,zhang17,CBMSHealth21}  & \cite{healthmonitor21} & \cite{healthmonitor21}\cite{Awasthi}\cite{Yan13,Amin15,zhang17,CBMSHealth21} & - & - & -\\
    WMSN & \cite{SecurityWMSN}\cite{HARCI21}\cite{Wanghealth21} & \cite{SecurityWMSN}\cite{Wanghealth21}\cite{xTSeH} & \cite{SecurityWMSN} & \cite{SecurityWMSN}\cite{Wanghealth21} & -& \cite{HARCI21}\cite{Wanghealth21}& -\\
    Blockchain Healthcare Systems & \cite{healthcare4.020}\cite{Griggs18}\cite{EHR19}& \cite{healthcare4.020}\cite{Griggs18}\cite{EHR19}& \cite{Griggs18}& \cite{Griggs18}& -& -& -\\
    
    \textbf{Smart Utility}\\
    Smart Gas Management System & -& -& -& -& -& \cite{smartgaskesh17}& \cite{smartgasshrestha19,smartgaskesh17, smartgasdong17,smartgasfeng19}\\
    Smart Water Systems & -& -& -& -& -& \cite{smartcitydriverwater21}& \cite{smartwatermudumbe15,smartwatercasestudy16,irrigation19}\\
    Smart Grid & \cite{smartgridhome}\cite{xiaetal12,wazid17,Srinivas21,grid20}& \cite{smartgridhome}\cite{xiaetal12,wazid17,Srinivas21,grid20}& \cite{IoTenergy20}& \cite{IoTenergy20}\cite{smartgridhome}\cite{wazid17}\cite{Srinivas21} & \cite{EI19}& \cite{xiaetal12}\cite{IoTenergy20}\cite{grid20}& -\\
    Blockchain Metering & \cite{blockchaingrid19}\cite{blockchaingridsurvey21}& \cite{blockchaingrid19}\cite{blockchaingridsurvey21}& \cite{blockchaingrid19}\cite{blockchaingridsurvey21}& \cite{blockchaingrid19}\cite{blockchaingridsurvey21}& -& -& -\\
    
    \textbf{Smart Transport}\\
    UAV & \cite{UAVcharging}& \cite{UAVcharging}& -& \cite{UAVcharging}& \cite{UAVcharging,AGMEN18,SAG18}& \cite{AGMEN18}\cite{SAG18}& -\\
    Smart Charging for EVs & \cite{vehiclecharging20} \cite{cpscharge}\cite{securityassement}& \cite{vehiclecharging20}\cite{cpscharge}\cite{securityassement}& \cite{cpscharge}& \cite{vehiclecharging20} & \cite{vehiclecharging20}\cite{cpscharge}\cite{securityassement}& -\\
    IoV Distributed Trust & \cite{inTransitVehicle20}\cite{conditionalanonymousiov21}& \cite{inTransitVehicle20} \cite{conditionalanonymousiov21}& -& \cite{conditionalanonymousiov21}& -& \cite{inTransitVehicle20}& -\\
    Traffic Management Systems & -& -& -& \cite{RSUOBU20}\cite{RSU21}& \cite{trafficontroldesign20}\cite{antcolony20}& \cite{trafficmonitoring20}\\
    
    \textbf{Smart Homes}\\

    Integration With Smart Grid & -& -& -& -& \cite{smartgridhome} \cite{Samie2019}& \cite{smartgridhome}\cite{Samie2019}\cite{homeenergygrid}& -\\
    Access Control In Smart Home & \cite{securekeysmarthome20}\cite{smarthomeauthentication20}& \cite{securekeysmarthome20}\cite{smarthomeauthentication20}& -& \cite{smarthomeauthentication20}& -& -& -\\
    Blockchain Home Systems & \cite{homechain20}\cite{SH-Block}\cite{blockchainhomedata}& \cite{homechain20}\cite{SH-Block}\cite{blockchainhomedata}& \cite{SH-Block}& \cite{homechain20}\cite{SH-Block}& -& -& -\\
    \bottomrule
    \end{tabular}}

    \caption{\textcolor{black}{Research topics and examples under different smart domains, group according to security and privacy sub-issues.}}
    \label{tab: examples}

    \end{table*}
    
    \subsection{Common Characteristics of Applications in Smart City}
    \textcolor{black}{In the preceding paragraphs, we have introduced the smart city landscape. Here, we summarize what a standard smart city application would look like and the characteristics that it will have.}
    
    \textcolor{black}{\textbf{User-centric:} Besides the common objectives such as optimizing resource utilization and increase system efficiency in conventional systems, improving user experience will be a main objective for standard smart city applications. There will be more focus on providing personalized service and designs that suit the preferences of the users.}
    
    \textcolor{black}{\textbf{Ubiquitous:} Smart city applications collect a vast volume of user data for insight extraction using ubiquitous sensors and devices located around the city. The data collected from these sensors may be shared between platforms and applications, and can be utilized to perform various data analysis before being transformed into actionable insights at the end devices.}
    
    \textcolor{black}{\textbf{Data-centered:} To enhance their operations, most smart city applications will leverage the big data landscape and take a data-driven approach. Better system automation and pattern prediction can be accomplished by employing data intelligence to increase both efficiency and user experience.
    }
    
    \textcolor{black}{\textbf{Interoperable:} In many smart city applications, we can expect devices with varying capabilities and resource constraints to participate in the same system, and there will be constant data exchange within the system and with external parties, imposing high interoperability requirements on the smart city application. For example, most of the applications relies on cloud storage directly or indirectly, for storing the massive data incurred.
    }
    
    \begin{table*}[ht!]
    \centering 
    \begin{tabular}{ |p{1.5cm}|p{7cm}|p{8cm}|  }
    \hline\hline
    \bf{Protocol} & \bf{Security Features} & \bf{Security Challenges} \\
    \hline\hline
    
    Zigbee & Zigbee Trust Center is responsible for key management and distribution in the network, and it provides cryptographic key establishment and frame protection. (AES 128-bit) & Zigbee is susceptible to jamming, traffic analysis attacks. Since different application layers on the same device trust each other, a compromise on one of the layers can escalate to the other layers. In addition, network keys can be extracted from the hardware, resulting in concerns for the physical capturing attack \cite{zigbeezwaveBLE}. \\
    \hline
    Z-wave & Z-wave network contains master and slave nodes. It provides authentication service, packet encryption, and integrity protection. (AES 128-bit, Message Authentication Code in Cipher Block Chaining mode) & Z-wave is susceptible to traffic analysis, spoofing, MITM attacks. Z-wave devices trust all MAC Protocol Data Unit, making them vulnerable to impersonation attacks \cite{zigbeezwaveBLE}. \\
    \hline
    Bluetooth Low Energy (BLE) & BLE provides encryption, authentication, message integrity checks, and 12-byte signature to prevent relay attacks in mode 2. In addition, BLE changes the address of devices frequently to reduce the ability for attackers to track. & BLE uses elliptic curve cryptography, which is not supported in legacy mode pairing. Thus, devices in legacy mode pairing are susceptible to MIMT and eavesdropping attacks \cite{zigbeezwaveBLE}. \\
    \hline
    IEEE 802.15.4 & The security of IEEE 802.15.4 is implemented at the MAC layer. Security features such as encryption and authentication, access control, confidentiality protection are provided. (AES)& IEEE 802.15.4 is susceptible to active tampering, jamming, eavesdropping attacks. Many attacks aim at different layers of the protocol \cite{coap}. \\
    \hline
    6LoWPAN & 6LoWPAN can operate over other platforms such as Ethernet, Wi-Fi, IEEE 802.15.4, and it has no security features. & 6LoWPAN does not provide any security feature and relies on other protocols or applications to complement its security requirement \cite{coap}. \\
    \hline
    Routing RPL & RPL provides message confidentiality and integrity. (AES, CCM, consistency check using challenge-response) & RPL is susceptible to internal attacks \cite{coap}. \\
    \hline
    CoAP & CoAP requires other protocols to provide it with security features. E.g. CoAP can rely on Datagram Transport Layer Security (DTLS) for encryption.	& Although DTLS is often implemented to provide security features for CoAP, DTLS is not fully compatible with CoAP (i.e., it does not suit CoAP proxy modes) \cite{coap}. \\
    \hline
    
    \end{tabular}
    \caption{Summary of the security features and challenges of wireless communication protocols for smart city IoT apps.}
    \label{tab: protocol}
    \end{table*}
    
    \subsection{Common Challenges of Applications in Smart City}
    Standard security requirements for smart city applications include confidentiality, integrity, availability, and accountability, as well as practical security considerations such as privacy, scalability, mobility, fault tolerance, interoperability and others. A detailed set of security requirements for smart city applications can be found in \cite{securityRequirementsIoT20} and also in \cite{20securityconsideration}.
    
    Aside from conventional security challenges such as maintaining data communication confidentiality, smart city applications will encounter a slew of new ones to meet those security requirements. For example, while security features in wireless communication technologies such as Wi-Fi have been extensively researched, many of the IoT wireless communication protocols have yet to receive adequate security updates. In Table \ref{tab: protocol}, we present the security properties of the most commonly used wireless communication protocols for smart city IoT applications, as well as the security challenges associated with them.
    
    \textcolor{black}{By examining the features of smart city applications in the preceding subsections, we can derive a set of security and privacy challenges that smart city applications will have to overcome. For example: }
    \begin{itemize}
    
    \item Securing wireless data communication between systems and between devices of varying capabilities in implementing security mechanisms is a significant challenge. Different parts of the wireless infrastructure in smart cities can be owned and operated by both the government and third-party service providers. Some parts of the wireless network can be vulnerable due to cross platform interactions and unclear security responsibilities. For example, in many TMIS applications, patient data in health monitoring systems is required to be transported from homes to government-controlled hospitals. The system will be vulnerable if neither party is concerned about the secure interaction between them or makes an invalid assumption about the effort that the other will place on securing the connection. Thus, integrated and end-to-end security solutions that meet the security requirements and wireless constraints, e.g., limited spectrum, are needed.
    
    \item Allowing heterogeneous devices to interact in a safe and efficient manner is another significant challenge. We expect constant interactions with a variety of public and commercial systems in the smart city environment, and many of these systems are vital for public security. Public infrastructure, such as the intelligent traffic control system and the emergence broadcast system, must be secure. Otherwise, attacks on the IoT systems can interfere with critical system components such as traffic lights or radio networks, resulting in severe physical and economic loss.
    
    \item The security requirements of data generated in smart city systems are time-dependent, and providing adequate protection for sensitive information throughout its life cycle is a challenge that smart city applications must overcome. For example, video footage in public surveillance systems exhibits high sensitivity due to privacy concerns, but the degree of sensitivity depends on the recency of the video data. The video footage today can disclose more important information about the status of inhabitants than the video footage taken three months ago, so the more current the data, the more security protection is required here. As a result, providing adequate protection throughout the data life cycle necessitates careful consideration, especially given the massive data scale.
    
    \item Smart city applications have to present a unified representation of the heterogeneous data acquired, which is both crucial and challenging. A unified representation of heterogeneous data types can support effective insight generation across devices and systems. It also allows systems to access the context and the security risk of the data exchange. Thereby, a more appropriate security setting can be established, avoiding unwanted interactions with sensitive data and preventing data loss. For example, with unified representation of data, smart health systems and smart home systems can communicate efficiently with unified data representations to extend the physical range of their services and notify security breaches to prevent escalation of attacks from one system to another.
    
    \item Another challenge for smart city application is to provide solutions for remote mass updates and system recovery in the event of an attack/intrusion. More malicious attempts by attackers probing for the weakest link in the system are likely to occur due to the large-scale physical deployment of IoT devices in smart cities, which increases the attack surface for attackers to interfere with system operations. At the same time, due to the massive deployment scale, a successful intrusion into these end devices might cause a tidal effect, compromising the security of the system and even paralyzing it. Consequentially, the effect of intrusion will be very significant and impacts many devices. For example, attacks on utility systems can paralyze the public infrastructure in cities within minutes. Thus, it is necessary to have solutions for efficient remote mass updates and attack recovery to increase system resilience.
        
    \end{itemize}

    In the following section, we will evaluate the specific context of each smart city domain and their security requirements from an ANT-centric perspective, presenting existing security proposals and some of their shortcomings

\section{Cybersecurity in Smart City Domains \label{sec: context}}

In this section, we will provide a systematic study of the security context in each smart city domain, one by one, beginning with an overview of developments in each smart domain and the characteristics that applications in each domain will have. Following that, we examine the security requirements of these applications from an ANT-centric perspective and give examples of existing security solutions that may meet such requirements.

\subsection{\textbf{Smart Healthcare}}
    \begin{figure*}[htb]
    \begin{center}
        \includegraphics[width=0.8\linewidth]{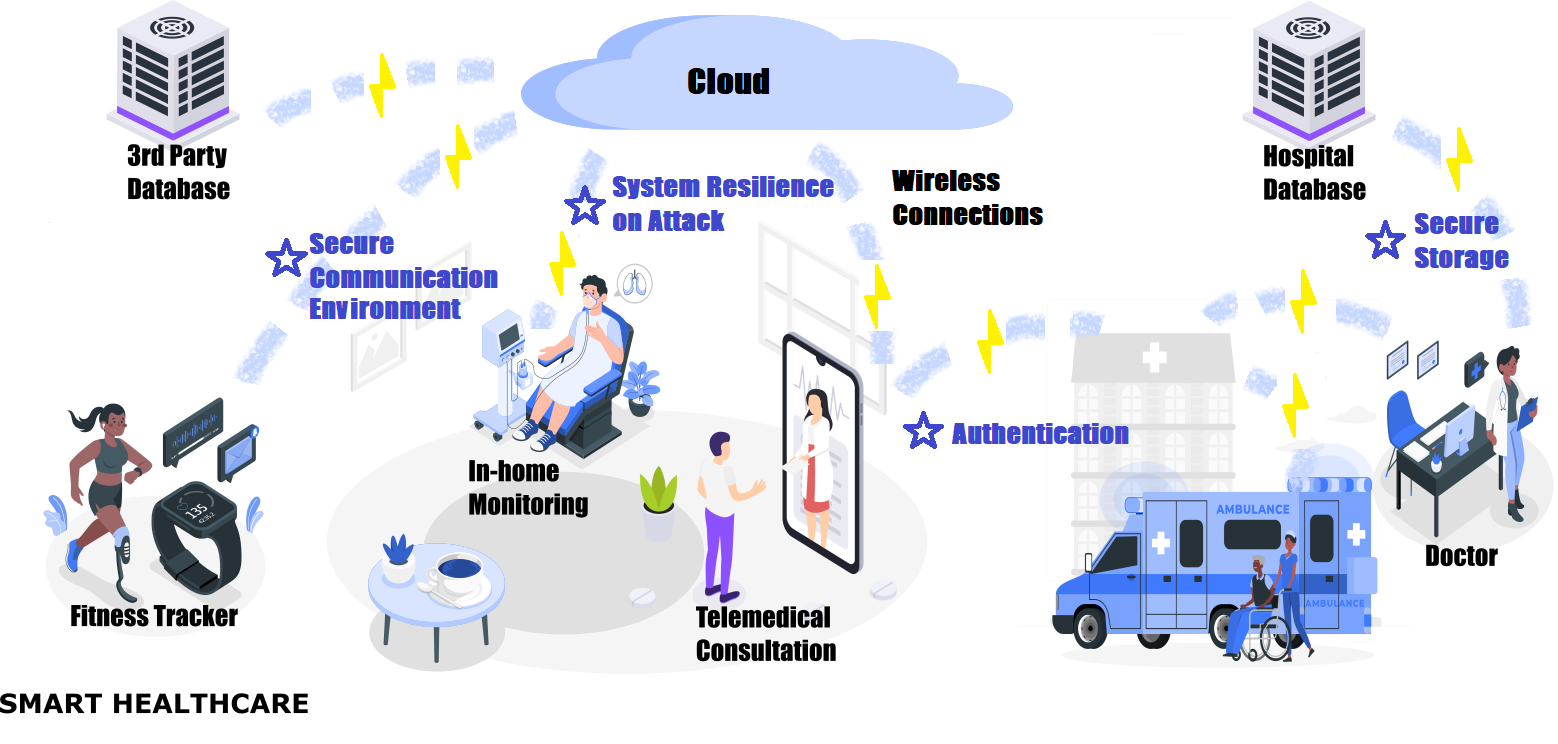}
        \caption{An illustration of major smart healthcare developments and some of the possible connections between them.}
        \label{fig:smart healthcare}
    \end{center}
    \end{figure*}
    
    \subsubsection{\textbf{Overview of Developments in Smart Healthcare}}
    Aging will soon be the major issue that many nations must address, especially in developed countries with a high average life expectancy. The need for elder care increases significantly, while the establishment of an affordable, qualified and responsive healthcare system remains of paramount importance for the quality of life of people. Fortunately, recent technology developments have paved the way for the transformation of the healthcare system into one that is data-driven, cost-effective, proactive, personalized and sensitive. 
    
    An illustration of major smart healthcare developments and some of the possible connections between them is shown in Figure \ref{fig:smart healthcare}. For example, Telecare Medical Information System (TMIS) is a telehealth system \cite{TMIS} that offers remote health monitoring and consultation services for its users. In TMIS, IoT sensors such as smart wearable body sensors, smart healthcare sensors and other IoT sensors are used to provide real-time or periodic updates on patient physiological signals. Information such as ECG data, physical movements, and air quality are collected and analyzed or monitored on remote servers for medical professionals to analyze medical conditions.
    
    Other developments for smart healthcare include Wireless Medical Sensor Network (WMSN) and Blockchain healthcare systems. WMSN is a network formed using wireless sensors deployed on the human body, which is responsible for monitoring vital health signals and transmitting data out of the network periodically \cite{abdulkarem2020wireless}. At the same time, blockchain healthcare systems present a distributive solution to protect the data integrity of personal health data in the cloud, utilizing the immutability of blockchain transactions.
    
    \subsubsection{\textbf{Characteristics of Applications}}Smart healthcare applications often exhibit the following characteristics:
    
    \textcolor{black}{\textbf{Data-driven:} Many of the proposed smart healthcare systems rely significantly on sensor data to perform functions such as health monitoring, medication reminders, and tele-healthcare consultations. These systems are highly data-driven in that their logic flow always centers around the use and storage of data. Depending on the nature and variation of the collected data, smart healthcare systems take their respective decision paths to trigger their procedures and alerts. Hence, the security design of smart healthcare systems will then mandate strong protection towards data flow and data interactions concerning sensitive patient data.
    }
    
    \textcolor{black}{\textbf{Cost-effective:} Smart healthcare systems are often built to be cost-effective and to achieve affordability to include more people who may not be able to enjoy healthcare services today. The current smart healthcare system mainly revolves around two types of smart healthcare IoT devices: (1) clinical healthcare devices; (2) personal healthcare devices. Examples of clinical healthcare devices include smart continuous monitors for heart rate and glucose level. Personal health devices, on the other hand, act as portable data collectors and aggregators to provide remote monitoring and health measurements for personal healthcare advice or care. They include devices such as smart wearable and self-monitoring activity trackers. With the use of such devices, many patients could choose home monitoring and recovery as well as remote medical consultation for minor illnesses, saving money on hospitalization and travel. Such developments will impose strict constraints on the accuracy and integrity of sensor readings and the availability of smart systems, requiring a robust and secure environment for data communication.
    }
    
    \textcolor{black}{\textbf{Proactive:} The conventional healthcare system is often passive and reactive, where patients seek medical advice only when severe symptoms appear. This can lead to the escalation of illnesses where early stage treatments are almost impossible when patients do not pay much attention to their health status. In smart healthcare systems, constant health monitoring is employed to identify critical health signals on a regular basis and to warn users of abnormal health statuses, allowing for early intervention of health conditions. Hence, the design of smart health care systems would extend their considerations to include monitoring pre-illness factors that can potentially contribute to illness.
    }
    
    \textcolor{black}{\textbf{Personalized:} Smart healthcare systems are highly personalized to cater to different healthcare needs of individuals. Depending on the user's health history and conditions, smart healthcare systems will have a different monitoring emphasis. Professional medicare will also have a set of features customized in smart healthcare systems to cater to various illnesses and diseases. Nevertheless, such personalization will necessitate a large database architecture in smart healthcare systems that can adjust the content of their services based on individual demands.
    }
    
    \textcolor{black}{\textbf{Sensitive:}} Healthcare data are very sensitive and heavily regulated, and applications that interact with it must provide extensive data protection. A successful attack on smart healthcare systems could result in life-threatening situations. Although the size of the deployment for these systems may not be as large as that of smart utility systems and the deployment environment is often in hospitals or residences with limited physical access, the hardware and software security of the components of the smart healthcare system is crucial. Meanwhile, patient anonymity and data confidentiality would impose heavy privacy concerns on such systems.

    \subsubsection{\textbf{ANT-centric Security Requirements}}
    In smart healthcare applications, it is vital to protect critical activities that communicate or store sensitive information about personal health information in their operations, as shown in Figure \ref{fig:smart healthcare}. Since cross-platform and cross-device data communication is expected in many smart healthcare applications, a secure communication environment is necessary because it is difficult to account for the security of all possible data interactions between devices. In addition, secure storage of sensitive medical data is also crucial for maintaining user privacy and data integrity.
    From a network perspective, it is crucial to ensure that all critical data communications occur at critical nodes that will involve critical data activities are authenticated to prevent illegal access and data confidentiality violations. For example, it is essential for both the patient and the doctor to authenticate themselves to the systems in order to establish mutual trust.
    From a things-centric perspective, we consider security issues that are closely tied to healthcare devices and systems. To avoid mishaps in smart healthcare applications that deal with sensitive health data, it is critical to have system resilience against attacks, where systems can detect suspicious activity that risks data integrity and confidentiality, especially when actual medical treatments are involved.
    
    Considering the characteristics of smart healthcare applications, these security requirements are among the most important criteria that these applications should meet. In this part of the section, we will explain each of the requirements in detail and present some of the related proposals that can be considered for meeting these requirements.

    \paragraph{\textbf{(Activity-centric) Secure Communication Environment}}
        While secure communication between system components is necessary, an end-to-end secure communication environment is also a prerequisite for smart healthcare applications. Since the security of a system is determined by how well the weakest link is protected, any parts of the system that are insecure can be the easiest target for attackers. In smart healthcare, our personal health monitoring devices, such as health trackers or smart watches, are constantly transmitting sensitive information about our physical conditions. These devices often have a limited ability to protect themselves and are transmitting important information, making them the most vulnerable components that deserve the most protection.
        
        For instance, suppose we constructed a complicated cryptographic technique to protect wireless communication between the third party database that stored our health data and the cloud and we did not establish a secure communication environment between the cloud and the end device. In that case, an attacker could instead attack this connection and capture all the important personal data passed. Thus, we should enhance the security of our smart healthcare applications by extending our security protection right from the start, from where data generation occurs. While there have been many proposals that could provide valuable insights towards fulfilling this requirement, most of them focus on securing parts of the system, such as end-to-end authentication \cite{JIANG2017182} or end-to-end secure key management \cite{ABDMEZIEM2015184}, rather than the complete end-to-end communication security. 
        
        In \cite{MOOSAVI2016108}, there is a proposal for end-to-end security by enforcing the adoption of the same security protocols throughout the system. The work in \cite{MOOSAVI2016108} is a secure architecture built upon the DTLS protocol for mobility enabled healthcare IoT. All parties communicating in the system mutually authenticate each other and establish secure sessions via a shared common key. Assuming that smart gateways have sufficient resources to perform security operations, smart gateways ease the computation burden on behalf of medical sensors by performing authentication and authorization of remote end-users. However, this application assumes that the application is implemented in smart homes/hospitals where authorized personnel cannot enter the physical premises, eliminating the need for authentication and authorization for remote healthcare centers or caregivers to those sensors. Thus, this design requires assumption of physical security of the environment, which may fail to protect against malicious insiders and physical intruders. On the other hand, SecureSense\cite{RAZA201740} combined the DTLS protocol with CoAP to achieve an end-to-end secure communication architecture for the cloud-connected IoT where certificate-based asymmetric cryptography is applied to compliment the crytographic capability of DTLS. Similarly, the work in \cite{GLISSA2019100} attempts to adapt the 6LoWPAN protocol to integrate lightweight end-to-end security to provide a secure communication environment, which they call 6LowPSec.
        
        Although a universal security protocol could improve system security by providing consistency, such protocols will need to specify further their context and the minimum requirement on the hardware and software capabilities for their implementation to improve their feasibility. Alternatively, we could have a centralized agency that defines a set of security protocols for different types of devices or data. Thus, any device wanting to communicate with each other must support the common set of security protocols and pass specific security checks before proceeding with the wireless communication.

     \paragraph{\textbf{(Activity-centric) Secure Storage}}
        The final stage of the data life cycle in a system will be its storage and management. A vulnerability in one of the wireless communication channels can expose a confined set of data exchanges to the attacker. When discovered early, its damage can be constrained within a relatively small range. On the other hand, a database vulnerability puts all information stored in the entire system at risk. As a result, in addition to improving data transmission security, ensuring data security at rest is also crucial, and it is especially so for sensitive healthcare records stored in cloud servers and hospitals.
        
        This paper classifies secure storage mechanisms into two types: (1) secure distributed storage (2) secure cloud storage. Distributed storage refers to the use of storage space on distributed nodes to provide storage services, avoiding a single point of failure. Kher and Kim \cite{securedistributedstorage05} proposed a set of security services that a secure storage system should support. It includes authentication and authorization, to validate the identity of the party accessing the database; availability, to ensure that the data is readily accessible for authorized parties, where the system would have to perform backup and recovery services; confidentiality and integrity, so that users can trust the system in keeping their data; key sharing and key management, to facilitate efficient sharing and retrieving of data; auditing and intrusion detection, to monitor and detect unauthorized access; usability, manageability, and performance, to provide smooth data storage services to enhance user experience. There has been an exploration on different technologies in achieving secure distributed storage, such as the combination of SHA-3 and Reed-Solomon (RS) erasure code \cite{SAONTRS18}, secure erasure codes \cite{secureerasurecode18}, Redundant Residue Number System (RRNS) \cite{CHERVYAKOV20191080}, blockchain \cite{blockchainstorage18}. Depending on system configurations and application context, some of these mechanisms are more suitable than others. For example, SHA-3 offers more compatibility than blockchain, where many existing systems have been built to support SHA-3.
    
        On the other hand, the authors in \cite{cloudstoragesurvey20} \cite{cloudstoragesurveytwo20} have surveyed the security issues for cloud storage. Issues such as data confidentiality, atomicity, data access and data breaches have been presented in these surveys, coupled with different security mechanisms to resolve those issues, such as identity-based encryption, attribute-based encryption, homomorphic encryption and searchable encryption. In smart healthcare applications, the location of data storage as well as the level of complexity in secure storage protections differ from one another. Nonetheless, system designers will need to determine the optimal trade-off between computational complexity and storage and access security, ensuring that data of varying sensitivity is adequately protected.
    
    \paragraph{\textbf{(Network-centric) Authentication}}
    In smart healthcare applications, authentication is the first line of defense for a secure communication channel. It validates the identity of the patient/user who is exchanging information with the medical systems. When paired with an authorization system, the authentication protocol could ensure that only authorized users/devices have access to or upload sensitive medical records, hence protecting confidential medical information from exploitation.
    
    Hardware-based authentication protocols for smart healthcare systems have been proposed in \cite{HARCI21}\cite{xTSeH}. The authors in \cite{HARCI21} proposed a lightweight authentication protocol, to mitigate the risk of physical node tampering and node replacement attacks on smart healthcare IoT devices. Their two-way two-stage authentication scheme implements the Physical Unclonable Functions (PUFs). The use of PUFs allows computationally expensive cryptographic operations to be performed in the PUF chips, saving precious resources on the smart device. While both logical and physical security during data exchange has been enhanced via the adoption of PUFs, the scheme requires the challenge and response pairs of every PUF to be stored in the cloud database, prior to the deployment of these devices and be retrieved for each round of authentication. This will limit the scalability of the scheme because when the number of smart devices in the network increases, there would be increasingly significant storage demand and time delay due to cloud access. 
    
    Similarly, xTSeH \cite{xTSeH} aims to build a trust chain from the hardware level to establish a network of trust for the smart healthcare IoT devices. The scheme extends the security functions of the Trusted Platform Module (TPM) to Smart Embedded Devices (SED, also refers to smart healthcare IoT devices here) that were not equipped with TPM, using TPM-chipped SEDs as the root of trust. Application-level proxies were designed as a communication bridge between the root of trust and non-TPM-chipped SEDs that requires authentication, verification, or trusted booting services. This scheme protects system integrity and ensures the authenticity of smart devices. However, these security services impose a significant delay on system performance, especially the validation of the non-TPMchipped SEDs. This is a challenging bottleneck because TPM chips process in serial, hence, further optimization will be needed.

    There are many proposals for software-based authentication as well. An overview of the different biometric-based authentication methods for smart healthcare can be found in \cite{HAMIDI2019434}. In \cite{Awasthi}, Awasthi and Srivastava proposed a three-factor authentication scheme for Telecare Medical Information System to enable secure data transfer between smart healthcare IoT devices and remote servers, as well as to provide medical personnel authorized access to patient data. This scheme is found to be vulnerable towards offline password guessing attacks, reflection attacks and fails to provide patient anonymity \cite{Tan3factor}\cite{Mishra14}. To overcome the drawback in Awasthi and Srivastava’s approach, both \cite{Tan3factor} and \cite{Yan13} suggested their version of three-factor authentication protocols using the smart card. Later, these schemes were found to be susceptible to offline password guessing attacks and lack patient anonymity as well \cite{Mishra14}. From that point, Mishra et al. \cite{Mishra14} proposed another three-factor authentication protocol using personalized smart card, which \cite{Amin15} \cite{zhang17} conclude to fail in protecting against server impersonation attacks, stolen smart card attacks, and being irresistible to MITM attack and replay attacks. The authors in \cite{CBMSHealth21} reviewed authentication protocols on the secure smart healthcare system adopting smart healthcare IoT devices. Furthermore, they proposed a mutual authentication scheme for the cloud-assisted medical information system, allowing the secure exchange of electronic medical records (EMR) and patient data without data duplication. However, they did not address the issue of secure communication during remote registration of medical devices and patients with the healthcare center, nor the secure storage of medical records in the system.

    \paragraph{\textbf{(Things-centric) System Resilience on Attacks}}
        Keeping resilience to deal with insider and outsider risks and attacks is crucial for smart healthcare applications. In health monitoring systems and wireless medical sensor networks, we have sensors and actuators that are always monitoring important health signals and transmitting this data to the cloud or hospital for health risk evaluations. If wrong information regarding a person's health is transmitted, incorrect medical assessments and threats can be made, perhaps leading to life-threatening consequences. As a result, it is critical for smart healthcare applications dealing with sensitive health data to detect suspicious activity that threatens data integrity and confidentiality, especially when actual medical treatments are involved.
        
        As described in \cite{rajamaki2016towards}, resilience mainly focuses on the integration of runtime situational awareness and prior risk analysis, which means that a resilient system should have the ability to take proactive defense and mitigate countermeasures against attacks to maintain system security and robustness. As one of the most important security posture monitoring tools, intrusion detection systems (IDS) can identify suspicious activities and detect possible attacks to make the network run in more security and normal operating state. There exist many mature intrusion detection techniques for the traditional network system. However, these IDS techniques can not be applied to IoT systems directly because of the many different characteristics between healthcare IoT and traditional network systems. Complex and high computational techniques are not suitable for healthcare IoT system as it is constrained, and there are many lightweight health trackers/sensors in smart healthcare applications, which we should consider their memory capacity, battery life, and network bandwidth when designing security detection approaches \cite{chaabouni2019network}. 
        
        New machine learning and deep learning methods for detecting abnormalities in smart healthcare applications have emerged as a result of recent breakthroughs in artificial intelligence. Such methods place little computational burden on the user device and can detect anomalies with astounding accuracy. In \cite{electronics10121375}, the authors developed an IDS for network traffic in smart healthcare systems by combining a genetic algorithm and the Random Forest algorithm on network traffic data. Their proposed algorithm achieved a detection rate of 98.81\% and a false alarm rate of 0.8\% on the NSL-KDD dataset. However, this approach can be further improved by considering the patient's biometric data as well. For example, authors in \cite{IDShealthcare} build their IDS utilizing both the network traffic data and patients’ biometric data. It has been discovered that by training with both biometric and network traffic data, the detection accuracy of IDS will be similar to, if not greater than, that of the best performing using either traffic or network data.

\subsection{\textbf{Smart Utility}}
    \begin{figure*}[htb]
    \begin{center}
        \includegraphics[width=0.8\linewidth]{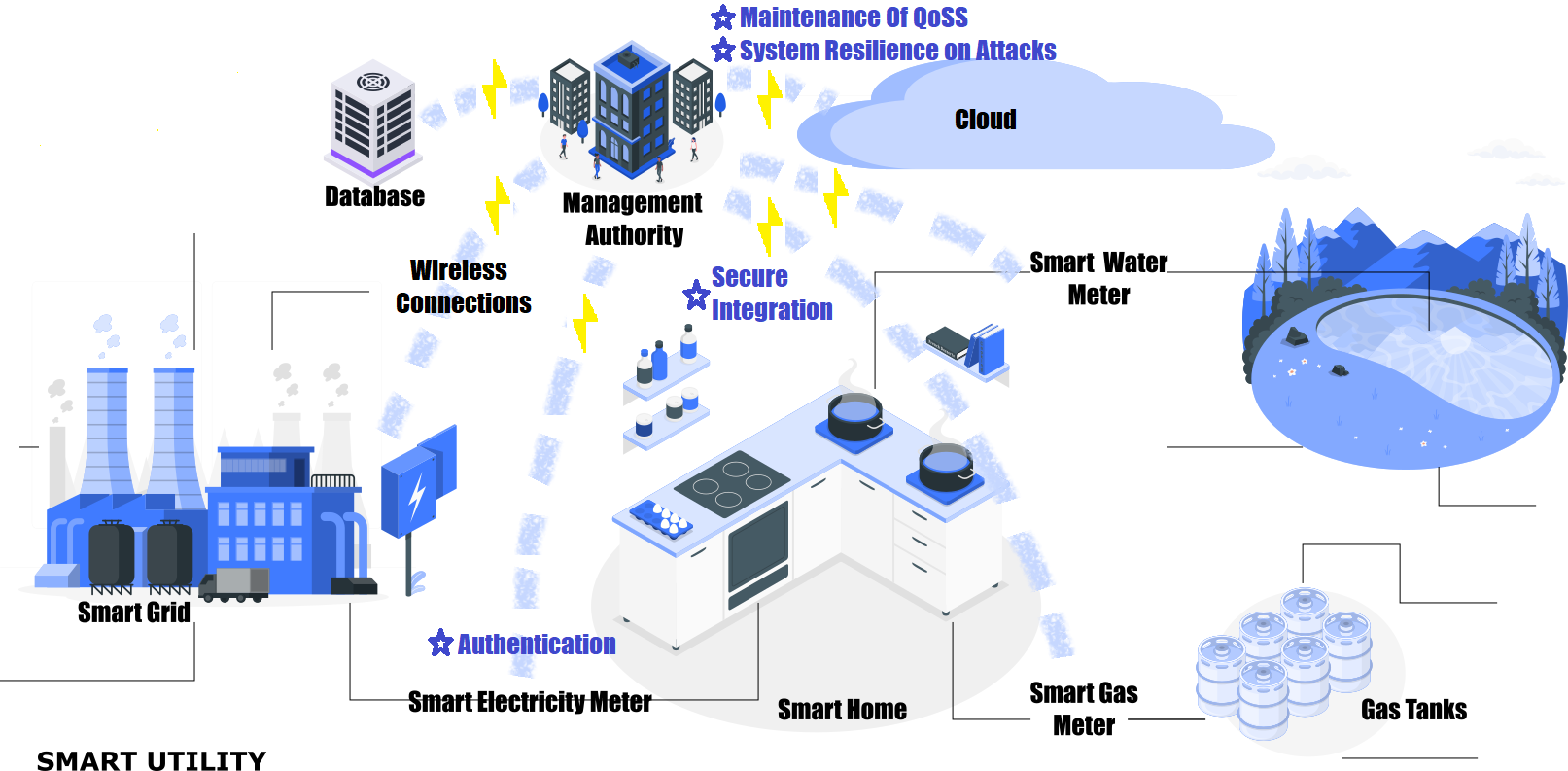}
        \caption{An illustration of major smart utility developments and some of the possible connections between them.}
        \label{fig:smart utility}
    \end{center}
    \end{figure*}

    \subsubsection{\textbf{Overview of Developments in Smart Utility}}
    Smart utility generally refers to public gas, electricity, and water systems that use smart infrastructure to provide safe, more efficient, and reliable utility services. Such a system is often data-driven and proactive, with real-time detection capabilities and system intelligence, as well as one that allows for constant user interaction. 
    
    An illustration of major smart utility developments and some of the possible connections between them is shown in Figure \ref{fig:smart utility}. There are three common types of smart utility systems: smart gas management systems, smart water systems and smart grids. Smart gas management systems incorporate smart sensors on our gas systems, mainly designed for system monitoring (load monitoring) and detection (gas or fire detection). Smart water systems enable cities to manage better and administer their water systems. Its main issue is the lack of consistent methodology and a unified data interpretation to develop a comprehensive water management infrastructure that encompasses all parts of a city’s water network, resulting in partial solutions that address specific problems in a given context. On the other hand, a smart grid transforms the electric power grid to allow interactive real-time connectivity between users and devices, allowing two-way high-speed communication across applications for smart system monitoring and control. A successful attack on the electrical grid has the strongest instant negative ramification among all attacks on public infrastructure.
    
    \subsubsection{\textbf{Characteristics of Applications}}Smart utility applications often exhibit the following characteristics:
    
    \textcolor{black}{\textbf{Data-driven:} IoT sensors are placed around critical infrastructures to constantly monitor the physical condition of the system and its components, providing important system measurements and feedback for performance analysis. For example, pressure sensors and temperature sensors are common IoT sensors deployed to monitor the physical state of a system component in smart utility systems. With such data collected, smart utility systems can generate useful insights on consumption patterns, system maintenance needs and make optimal operational decisions. Hence, smart utility system designs are often centered around data collection, processing and storage.
    }
    
    \textcolor{black}{\textbf{Proactive:} In traditional utility systems, we regularly perform periodic maintenance services to monitor system condition and prevent unanticipated system breakdowns. This is time consuming and inefficient since we are unable to fully utilize our resources; yet it is needed to avoid unanticipated service disruptions. In the event of a breakdown, we undertake reactive repairs to save the faulty system components. With smart utility systems, we are able to constantly monitor our system components and even predict the optimal maintenance timing from the data generated. By taking a proactive approach, the smart utility system can optimize resource utilization and deliver better service quality. To achieve proactiveness, smart utility systems often rely on strong machine learning models and data analysis techniques to predict patterns and provide actionable insight, which further exaggerates the importance of data integrity.
    }
    
    \textcolor{black}{\textbf{Real-time detection:} Sensors for utility systems have a higher requirement for battery life than in other applications. Many of these sensors will be placed in physically inaccessible locations, such as in the underground tunnel or bottomless water tanks constructed miles deep in the ground. Replenishing the batteries inside these sensors in such scenarios would be a challenging task, and hence it is often left unmonitored. Thus, real-time detection is a special feature that smart utility systems can afford with the help of low-energy-consumption IoT devices and wireless communication protocols.
    }
    
    \textcolor{black}{\textbf{User participation:} A key functionality of smart utility systems is the interactivity between parties. By leveraging on the insight generated from system data, better market decisions can be made by utility companies, energy traders, and even consumers to balance the demand and supply of utilities with market force. Since more data communication between different parties is expected, strong authentication is a prerequisite for reliable user participation.
    }

    \subsubsection{\textbf{ANT-centric Security Requirements}}
    From an activity-centric view, securing essential activities that may affect the availability and accuracy of data communication over the full data life cycle is critical to the security of the system in smart utility applications. Integrating data collected from numerous smart utility sensors, applications, and processes is a critical and complicated task, as illustrated in Figure \ref{fig:smart utility}. As a result, in order to preserve the overall security of the ecosystem, we must address the potential security concerns associated with such integration. Moreover, smart utility applications will require efforts to maintain Quality-of-Services in order to improve system performance and user interactions.
    From a network-centric perspective, device authentication is an essential security requirement against network attacks, ensuring that all critical data communications occur at critical nodes are appropriately protected so that data integrity and system availability can be maintained.
    From a things-centric perspective, we consider security issues that have a direct impact on smart utility devices and systems. Due to the extensive attack surface of smart utility applications and the strong instant negative ramification on public infrastructure, there is a requirement for system resilience on attack in the form of a strong healing ability after attacks occur to limit the damage caused by attacks.
    
    Considering the characteristics of smart utility applications, these security requirements are among the most important criteria that these applications should meet. In this part of the section, we will explain each of the requirements in detail and present some of the related proposals that can be considered for meeting these requirements.

    \paragraph{\textbf{(Activity-centric) Secure Integration}}
        Integration for data collected from various sensors, applications, and processes is challenging. It is difficult to group multiple technologies into one for smooth data collection. The context and environment from which the data value is collected vary significantly. The smart grid may employ blockchain for autonomous user transactions on electricity, where electricity is priced in terms of system currencies, whereas smart water management systems price based on the cost of water treatment per household and real consumption. This contributes to barriers in providing consistent data interpretation, and additional contextual data need to be present to generate useful insights for end users.
        
        More importantly, we must consider the potential security risks associated with such integration. Since mandating a uniform set of protocols to be implemented by different utility system is currently challenging, the systems will try to connect resource-constrained devices securely using the methods that those devices support \cite{FadhilKhalid_YAmeen_2021}. Some devices can support a basic mechanism, while others do not attempt to protect user information. For example, a sophisticated smart utility management server may protect consumption data by encrypting it with strong cryptographic mechanisms; however, these data are decrypted by smart gateways and distributed to end point devices such as smart meters installed in homes, where the data is not protected. Thus, these devices became the weakest link in the system, where attackers could easily breach in, and attacks can rapidly escalate to the rest of the system. Hence, efforts for secure integration would be necessary to enhance system security.
        
        This criteria can be fulfilled in a variety of ways. Although there are no proposals for secure integration of smart grid, smart water, and smart gas management systems, we can learn valuable insights from system integration proposals. For example, the authors in \cite{FadhilKhalid_YAmeen_2021} have provided a discussion on the recent effort towards secure integration. Similarly, a narrower integration perspective between IoT and cloud computing \cite{STERGIOU2018964}, blockchain \cite{blockchainiotintegration18} has also been studied.
    
    \paragraph{\textbf{(Activity-centric) Maintenance of Quality of Security Service}}
        Similar to Quality of service (QoS), Quality of security service (QoSS) is the concept that quantifies the level of security in a system, where a variable level of security services and requirements exist. The level of security in a system would then refer to the degree of security factors considered. This is a critical metric for ensuring the safe deployment of large-scale IoT applications in smart cities. Similarly, QoSS can enhance system performance and achieve higher user satisfaction to provide users or network tasks with a range of appropriate security choices \cite{QoSScynthia01}. 
        
        Smart utility systems are often large-scale networks with a wide range of target users. It is anticipated that more interactions between systems and users will happen to optimize resource utilization, such as energy trading in smart grid systems and consumption monitoring in smart metering infrastructures. While users are becoming more aware of the importance of their privacy and data, common users cannot judge the level of security in an smart utility system due to a lack of measurable security metrics. Even when a manufacturer gives assurances about the level of security that users can anticipate from their system, there is no way for users to verify that these promises are kept. As a result, there is minimal transparency for consumer checks and balances, making it difficult to build trust with the users in such situations.
    
        The development of QoSS parameters not only can build trust between users and manufacturers, but it could also gain user involvement in monitoring the system status to provide more responsive feedback on possible attacks. This is especially important for large-scale systems like smart utility systems, where it may take a long time before they can filter out any faults in the systems. As a starting point, the authors in \cite{SAVOLA201378} have provided a quantitative Expert Opinion Survey (EOS) on the choice of quality criteria for the security metric, evaluating possible factors to consider for QoSS. Factors such as accuracy, time dependability of the system, are possible candidates for the security metric.
    
        On the other hand, the method used in \cite{Anupam20} can be adopted for a data-centric approach for the development of QoSS. In this paper, the appropriate level of security that the data deserves is determined by the path that the data has traversed in the system. To determine the final security level of the data, one has to refer to the security level given to all the physical components that it has visited, the various source data used in formulating it, the context where the source data is generated. By considering the sensitivity of all nodes and operations related to the data, we could determine the importance of the data and thus define a suitable level of security to enforce in our system to protect it. This approach can be used as a starting gear for system designers to evaluate the level of security their systems need. In summary, maintaining QoSS through systems can help develop confidence between parties while also allowing for a thorough security evaluation for system designers to identify security risks early. This will encourage user interactions and better system monitoring via user feedback.

    \paragraph{\textbf{(Network-centric) Authentication}}
    Most of the existing literature on authentication of smart utility systems focuses on the smart grid, hence we will use the smart grid as an example to introduce how authentication can be done in smart utility systems. Due to the intrinsic scale of modern cities’ electrical grids, which are continuously connected to a huge number of devices and systems, there is a large “attack surface” for the attackers to exploit on \cite{smartcitydriverwater21}. This “attack surface” includes smart IoT devices, central servers, user end-point systems and all other possible entry points for the attacker to intrude the system. Furthermore, the smart grid’s significant reliance on communication and networking infrastructure exacerbates this problem, posing system security and privacy risks via the communication channel. 
    
    A successful attack on the electrical grid has the strongest instant negative ramification among all attacks on public infrastructure, as the majority of the city’s public systems will be paralyzed instantly. As a result, it’s critical to incorporate security and privacy considerations into smart grid architecture, to protect the IoT devices and the system components. Xia et al. \cite{xiaetal12} has proposed a simple and highly efficient secure key management system, with computation costs of 5ms and 4ms for a single authentication instance at the smart electricity meter and utility service provider respectively. However, \cite{wazid17} point out that the scheme fails to provide security features such as strong credential privacy and security for the session key. Likewise, \cite{Srinivas21} presented the AAS-IoTSG, a novel authentication scheme for the smart grid using ECC-based Schnorr's signature mechanism. Although it also comes with a considerable computation overhead, elliptic curve cryptographic (ECC) is much stronger than the conventional AES and RSA keys with the same key sizes in the current context. It is thus a more efficient candidate if we choose to employ cryptographic algorithms. For example, we choose a non-singular elliptic curve ${E_{q}(u,v)}$ of the form $y^2=x^3+u^x+v(modq)$ over a prime (finite) field ${Z_{q} ={0,1,…,q-1}}$ with a base point P. A 256-bit ECC key over this ECC curve is as strong as a 3072-bit RSA key. Hence, ECC can be a more suitable candidate to keep up with the growing cryptographic capabilities of the attacker.
        

    \paragraph{\textbf{(Things-centric) System Resilience on Attacks}}
        System resilience against attack from insiders and outsiders is crucial for smart utility. As described in \cite{rajamaki2016towards}, resilience mainly focuses on the integration of runtime situational awareness and prior risk analysis, and once it is attacked, it can also return to a normal and healthy operating state as soon as possible. This is especially important for smart utility applications where a slight disruption of utility services will result in an extensive impact on the operations of smart cities.
    
        Due to the limited memory capacity of the lightweight devices/sensors, IoT system such as the smart utility system is much easier to attack by the distributed denial of service (DDoS) attack. Besides, the basic components of an IoT system that generate data and make wireless communications are small and simple devices/sensors instead of human or complex machines. It becomes more difficult to secure data privacy in different smart city applications than traditional internet applications. For example, DDoS attacks on smart grid systems may cause power interruption, which will affect the normal life of individual users and hinder the production work of enterprises. Since most of the smart utility operations are time-critical and safety-critical, security mitigation countermeasures should consider meeting the critical conditions while devices/sensors or applications are under attack. For some applications that need to keep working in real-time, multiple security responses and recovery methods should be arranged to enhance the system's resilience. Due to the complexity of the smart utility ecosystem, where a large amount of data traffic is exchanged constantly between heterogeneous devices, designing for system resilience to attacks can be a very challenging security requirement but a necessary future direction before the adoption of smart utility systems.

 \subsection{\textbf{Smart Transportation}}
    \begin{figure*}[htb]
    \begin{center}
        \includegraphics[width=0.8\linewidth]{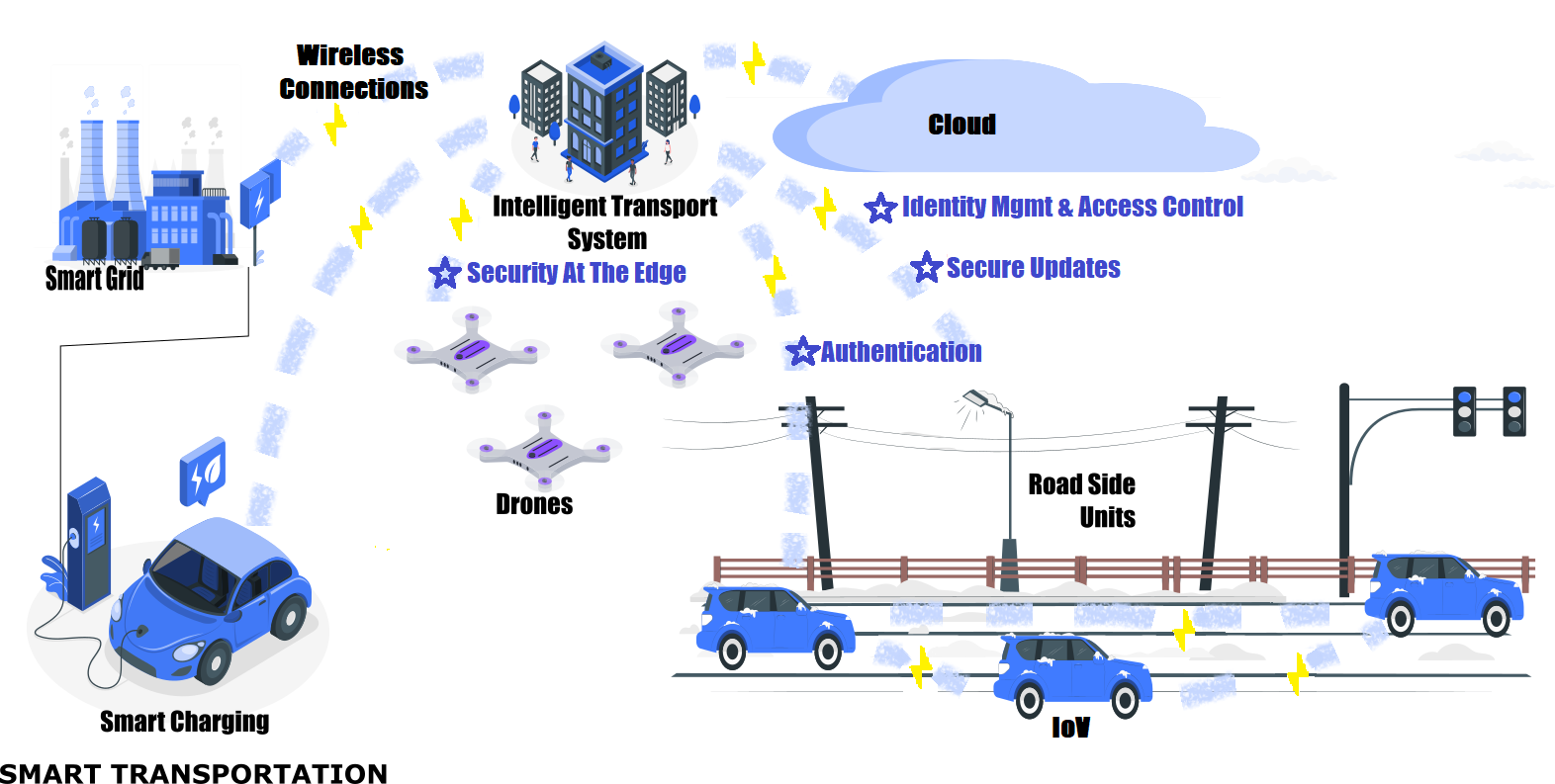}
        \caption{An illustration of major smart transportation developments and some of the possible connections between them.}
        \label{fig:smart transportation}
    \end{center}
    \end{figure*}
    
    \subsubsection{\textbf{Overview of Developments in Smart Transportation}}
    Intelligence transport systems (ITS) aims to improve the transport systems' road safety, efficiency, reliability, and user experience for all road users, including pedestrians, train riders, motorcyclists, and everyone else who participates in the transportation network of the society. As a result, a variety of smart transportation applications have evolved to transform our safety-critical transportation system into one that is data-driven, interoperable, and capable of providing real-time intelligence. An illustration of major smart transportation developments and some of the possible connections between them is shown in Figure \ref{fig:smart transportation}.
    
    Besides Unmanned Aerial Vehicles (UAVs) and smart charging for Electric Vehicles (EVs), Some of the main developments towards smart transportation include Internet-Of-Vehicle (IoV) distributed trust and intelligent traffic management systems. Internet-Of-Vehicle (IoV) distributed trust is an alternative trust management methodology compared to conventional centralized trust management where a single trust authority is responsible for generating authentication materials. Due to the extensive connectivity, changing topology, and requirement for privacy protection while guaranteeing proper traceability, authenticating IoV wireless communications is problematic. As a result, distributed trust management will aid in alleviating the present limits on IoV authentication tasks.
    
    On the other hand, intelligent traffic Management Systems includes smart traffic monitoring systems, smart road side units, and many more, they provide functionalities such as in-transit attestation, authentication, and support many other operations such as smart RSU, smart traffic control systems, smart traffic monitoring systems using UAVs to regulate and ensure safe driving practices, video-based traffic control for routing optimization using ant colony optimization, and vehicle-to-vehicle wireless communication.
    
    \subsubsection{\textbf{Characteristics of Applications}}Smart transportation applications often exhibit the following characteristics:
     
     \textcolor{black}{\textbf{Data-driven:} Real-time traffic data collection is critical for strategic transportation planning. With more information, we can make better informed decisions regarding our transportation plans and optimize our resource use. Sensors and cameras on vehicles and roadside infrastructure, provide vital traffic data such as vehicle density, travel speed and location. This data is transformed into useful information for tasks such as traffic forecasting, route planning, surveillance, and more, where we utilize it for generating transport intelligence. Given the criticality of transportation systems in public security, there is a strong need to ensure data integrity and availability for data-driven smart transportation applications.}
     
     \textcolor{black}{\textbf{Real-time intelligence:} In traditional transportation systems, traffic authorities typically invest significant resources monitoring current traffic conditions and attempting to resolve ongoing traffic accidents as quickly as possible to avoid traffic congestion. However, because of the massive scale of transportation networks, there is often a lack of manpower and technical capabilities to examine every single detail generated in the system. In smart transportation systems, real-time processing and insight generation can be done with the substantial computing power in the system and provide immediate feedback on the traffic condition, with a strong emphasis on resilience towards attacks on the availability of data and systems.}
     
     \textcolor{black}{\textbf{Interoperability:} Higher interoperability is expected in smart transportation systems where cross-platform interactions can be integrated to improve the resource utilization in smart cities further. For example, smart transportation systems are typically paired with the smart grid system to enable routing and charging optimization for electric vehicles. Similarly, end-user entertainment products such as the smart phone and smart watch can interface with smart vehicles to provide user-friendly transitions between activities. Thus, secure communication for smart transportation applications needs to also consider interactions across platforms and devices.}
     
     \textcolor{black}{\textbf{Safety-critical:} Both traditional and smart transportation systems prioritize safety, yet the latter can reach higher levels of safety than the former. In fact, one of the main reasons for implementing smart transportation systems is to improve traffic safety and decrease the occurrence of accidents caused by human error. With improved sensing and computing capabilities, autonomous driving could provide a safe and efficient alternative mode of transportation for metropolitan commuters. Nonetheless, the security and privacy of the data collected are the top concerns for implementing reliable and safety-critical smart transportation applications due to their implication for public security.
     }
    
    \subsubsection{\textbf{ANT-centric Security Requirements}}
    From an activity-centric view, securing critical activities that may affect the availability, integrity, and confidentiality of data communication over the full data life cycle is critical to the security of smart transportation applications. With reference to Figure \ref{fig:smart transportation}, there are various types of smart vehicles that exhibit high computation power and are where much of the crucial data processing and storing occurs. Thus, securing these edge devices is important for protecting critical information flow and data activities in smart transportation applications.
    From a network-centric perspective, vehicle and device authentication is a critical security requirement for establishing a trusted traffic network, guaranteeing that devices are who they claim to be. Furthermore, authentication can be integrated with identity management and access control to ensure that all important traffic communications are legal and adhere to the transportation authority's access restrictions.
    From a things-centric perspective, we consider security issues of components and vehicles in smart transportation infrastructures. Secure updates are especially crucial for smart transportation applications because of the vast network of public transportation infrastructure, where on-site maintenance is impossible.
    
    Considering the characteristics of smart transportation applications, these security requirements are among the most important criteria that these applications should meet. In this part of the section, we will explain each of the requirements in detail and present some of the related proposals that can be considered for meeting these requirements.

    \paragraph{\textbf{(Activity-centric) Security at the Edge}}
        Edge computing provides computation and storage offloading for many smart city applications. It uses edge devices to operate as microservers, collecting data from end nodes, aggregating and filtering it, providing real-time feedback and completing any user requests it can, and only sending intensive processing and permanent storage to the central server. Securing wireless communication at the edge is an essential task for reliable smart transportation applications. For example, UAVs can aid in edge network connectivity \cite{dai2020uav}, caching, and computation. However, UAV-based applications have not been able to widely apply for real-time and scalable tasks, due to their limited flight time, the requirement of periodic charging, and the lack of secure wireless communication protocols. Similarly, smart charging for Electric Vehicles (EVs) requires secure wireless communication between charging stations, service providers and EVs, where they are seamlessly connected to make charging-related decisions.
        
        In \cite{edgecomputingsurvey19}, the authors believe that the introduction of edge computing has increased the attack surface due to:
        \begin{itemize}
        \item weaker computation power of the edge devices, which are more vulnerable to existing attacks which are ineffective on most advanced systems now, such as small scale DDoS; 
        \item attack unawareness of the users, since users are generally unaware of the system states of sensors and panels if no user interface is provided; 
        \item OS and protocol heterogeneity, since there is little standardization across different edge devices; 
        \item coarse-grained access control, besides conventionally access permissions such as read and write, edge devices provide more capabilities that need to be regulated.
        \end{itemize}
        
        Besides the difficulties mentioned above, we have yet to establish a boundary for edge devices' responsibility. For example, in cases where devices are incapable of applying any security measures, we need to decide if the edge server should be responsible for encrypting the data. If the data is worth protecting, another challenge is how we, or the edge server, determine the security level that should be applied to it or treat all data equally. From the standpoint of data gathering, we must decide whether the edge server should listen to any device or only a restricted group of devices in the physical environment. As a result, maintaining security at the edge is a major requirement for smart transportation systems where vehicles are not centrally protected. This is especially true in the case of IoV, where a single vehicle can potentially disrupt the entire system's operation. 
        
        Different combinations of technologies can be used to satisfy this requirement. System designers have to consider the different interactions with edge devices and employ appropriate security mechanisms. For example, multi-factor authentication can prevent unauthorized access to edge devices, while complex cryptographic algorithms can secure communication between the cloud and the edge servers \cite{ROMAN2018680} \cite{abbas2018} \cite{edgecomputingsurvey19} \cite{ranaweera21} \cite{ranaweera21}.

    \paragraph{\textbf{(Network-centric) Authentication}}
        IoV is a complex IoT device capable of sensing and controlling other subcomponents for autonomous navigation, making driving safer and more enjoyable. By connecting wirelessly to roadside units and analytical cloud, its cognitive abilities are supported by an extensive smart transportation infrastructure that constantly provides traffic and environmental data. These smart vehicles are also expected to broadcast messages to the transportation infrastructure and vehicles in their immediate vicinity on a continuous basis to facilitate traffic information exchanges. While efficient data exchange is necessary for a pleasant driving experience, there is a primary requirement to guarantee the data authenticity; any malicious information could have disastrous effects on road safety. However, due to the extensive connectivity and rapidly changing topology caused by high mobility, authentication in the IoV has become very challenging. Furthermore, in order to protect users’ privacy, it is undesirable to authenticate IoV communications directly based on vehicle identities, because an attacker can simply infer the location of the vehicle by collecting broadcasted IoV messages. In order to protect user privacy, assigning pseudonyms to IoV that are frequently updated to ensure anonymity is a common way to establish conditional anonymous authentication. Such a method necessitates frequent changes to pseudonyms and is vulnerable to Sybil attacks, in which a single vehicle is capable of impersonating several vehicles using multiple pseudonyms. 

        As a result, to reduce the risk of Sybil attacks, it is necessary to be able to determine whether several signatures on the same message belong to the same vehicle, which leads to the concept of message linkable group signatures. However, there is a key security flaw in this scheme: it assumes that the group manager can be trusted. For strict adherence to the ‘zero trust principle’ where no party should be trusted, the author in \cite{conditionalanonymousiov21} proposed a novel Message Linkable Group Signature scheme with Decentralization Tracing (MLGSDT) for conditional anonymous authentication with abuse-resistant tracing in IoV. By combining the MLGSDT scheme and the decentralized group model, the proposed mechanism could achieve a distributed trust for IoV in smart transportation.
        
        Similarly, the work in \cite{vehiclecharging20} mainly concerned with authentication for electric charging between the ground stations and vehicles. The authors conducted a thorough analysis of the literature concentrating on the security and privacy issues of EV-related applications, hence, proposing a security mechanism to optimize EV charging station allocation in a safe and privacy-preserving way via lattice cryptography and authentication. In the proposed scheme, EVs broadcast their charging requests to aggregators near charging stations, which then communicate the request to an operator who assigns EVs to the charging stations. Finally, when an EV arrives at its designated charging station, it exchanges data with the charging station and stores it. Lattice signcryption, which performs digital signature and encryption concurrently to decrease time consumption and withstand quantum attacks, protects the data exchange. This scheme requires the initial registration of EV and the aggregator devices with the operator, where the operator is responsible for securely distributing the key pairs corresponding to their ID. However, because the system must keep separate pairs of public and private keys for each aggregator and EVs, the system cannot accomplish end-to-end security without considering key management and safe storage of the credentials utilized. Furthermore, the user's current trajectory, road conditions, and ride destination are not considered in the allocation, which might result in an assignment that is off the track for users, reducing the scheme's usefulness.

    \paragraph{\textbf{(Network-centric) Identity Management and Access Control}}
        Identity management and access control refer to the technology that makes a set of rules to define who meets the requirements for accessing, using, or modifying certain resources or data in specific devices of a system. Identity management and access control can also be called identity and access management (IAM). IAM ensures the security and productivity of the system by allocating the right resources to the right objects. It is a key technology to protect information security and avoid unauthorized data access by authorizing and revoking authorization based on delegation policies \cite{qiu2020survey}. In smart transportation systems, traffic data is both safety-critical and time-critical. Traffic authorities require an effective and highly secure IAM in order to deliver important traffic information, claim liabilities for traffic accidents and track violations of traffic rules.
        
        Unlike traditional Internet systems, IAM in smart transportation focus more on the digital management of devices that hardly require direct manual operation and involve the management and authentication of users and services. For example, the authors in \cite{7317862} believes that designing a security infrastructure, such as a Vehicular Public Key Infrastructure (VPKI), will enable secure conditionally anonymous vehicular communication. However, there are several challenges for implementing such IAM in smart transportation systems, including but not limited to revocation, credential abuse, default password risk, virtual eavesdropping, and some extreme environments for on-site access control management. More recently, there have been attempts to build IAM for smart transportation via the blockchain. By utilizing decentralization and immuntability of blockchain technology, blockchain-based decentralized identity management system can provide strong security and transparency for all parties in smart transportation systems \cite{STOCKBURGER2021100014}. However, there have always been computation and compatibility concerns for the adoption of blockchain technologies.
    
    \paragraph{\textbf{(Things-centric) Secure Updates}}
        Due to the enormous number of smart transportation infrastructures deployed, on-site maintenance is no longer feasible, especially for urgent mass updates for security patching. This has directed our interest towards remote updating mechanisms, performing updates and integrity attestation over the air. However, due to the wireless nature of such mechanisms, security concerns have become growingly prominent.
        
        A combination of different security approaches can be adopted to satisfy this requirement. For example, in \cite{SCUBA}, SCUBA (Secure Code Update By Attestation) is presented for resource-constrained IoT sensor networks to perform detection and automatic recovery of compromised nodes via secure code updates. SCUBA identifies compromised regions using authentication and software-based attestation and transmits the code update to repair the infected region. 
        Similarly, in \cite{pose10}, the authors proposed a Proof of Secure Erasure (PoSE) scheme for the secure remote update. In PoSE, the prover first performs secure erasure of all writable memory contents to remove possible malicious code before retrieving and installing code updates. However, it is pointed out by \cite{assured18} that SCUBA and PoSE have significant drawbacks that make them unrealistic for deployment. The assumption of strict local wireless communication, or the lack of support for version roll-back if the current version fails, contributes to their unsuitability. 

\subsection{\textbf{Smart Homes}}
    \begin{figure*}[htb]
    \begin{center}
        \includegraphics[width=0.8\linewidth]{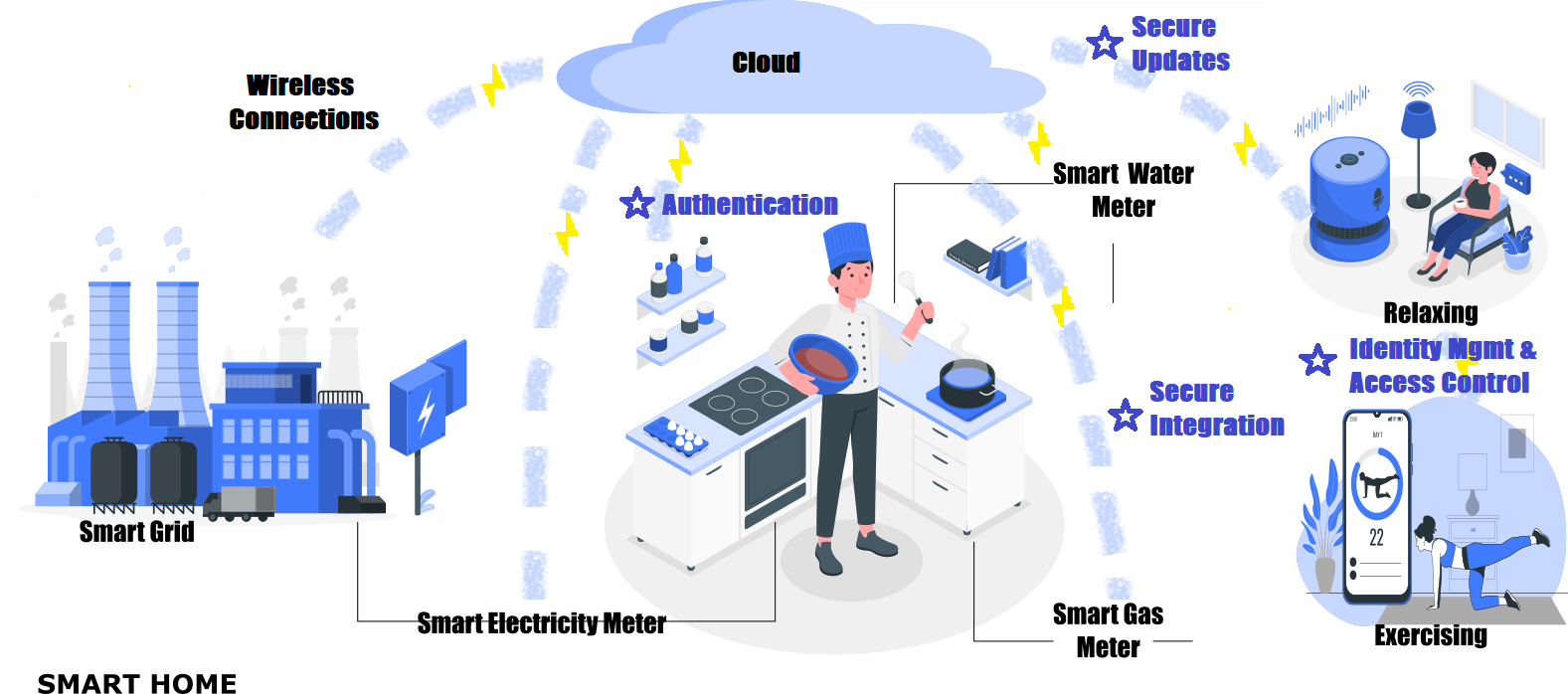}
        \caption{An illustration of major smart homes developments and some of the possible connections between them.}
        \label{fig:smart home}
    \end{center}
    \end{figure*}
    
    \subsubsection{\textbf{Overview of Developments in Smart Homes}}
    A smart home, also known as an automated home, is a home that is filled with the physical embodiment of an integrated and automated context-aware support system that improves our quality of life \cite{CASAS13}. The intelligence of our interactive home system arises from its ability to be context-aware. It gains its perception via sensor readings obtained from IoT devices, which quantifies the physical phenomenon emerging from human activities, to convey the context around the inhabitant \cite{smarthomeconcept06}. In summary, a smart home is one that is data-driven, context-aware, personalized and interoperable. An illustration of major smart home developments and some of the possible connections between them is shown in Figure \ref{fig:smart home}.
    One of the most significant advancement in smart home systems is their integration with the smart grid, enabling improved automation in smart homes by allowing various smart home applications to interact with the smart grid to obtain data such as utility consumption and market price. However, implementing this integration will necessitate cross-platform data communication and impose severe data security challenges.
    
    \subsubsection{\textbf{Characteristics of Applications}}Smart home applications often exhibit the following characteristics:
     
     \textcolor{black}{\textbf{Data-driven:} A smart home system is always data-driven and requires constant sensing and actuating to improve the user experience. The smart home typically consists of smart IoT devices equipped with a wireless communication interface, which forms a wireless sensor network. A central hub device is present to perform local processing, storage, data aggregation, and forward/receive information over the Internet. The cloud is expected to manage insight generation from the continuous, heterogeneous pool of data given by smart homes. Thus, the entire system flow of smart home systems is centered around the collection, processing, and storage of data. }
     
     \textcolor{black}{\textbf{Context-aware:} Context awareness, or the ability to gather information about the surrounding environment and adapt system behavior, is a fundamental feature of many smart home systems. Smart home systems can recognize events in the environment and forecast future patterns of interactions using sensor data, delivering relevant knowledge that allows systems to adapt to the lifestyle of the occupants. The capacity to deliver context-aware insights about the user raises serious privacy and security concerns concerning the use and storage of sensitive personal data.
     }
     
    \textcolor{black}{\textbf{Personalized:} From the contextual data, our smart home systems could deduce the current status and provide adaptive services as defined by the user or its default algorithm. A smart temperature sensor, for example, can detect the presence of human activity by comparing records in its database with a sudden spike in temperature, which can activate the user's default lighting. As a result, consumers can experience a personalized set of services in smart homes that adapt to their own behaviors and preferences. Nevertheless, such personalization will require model training over a large amount of sensitive user data and exhibit a strong need for data integrity and confidentiality.}
     
    \textcolor{black}{\textbf{Interoperability:} A smart home must be highly interoperable with cloud services, third-party applications, and consumer devices. In a typical smart home, the cloud maintains data security and secrecy and provides reliable services to enable the operation of third-party applications. These third-party applications are designed to provide interactive home automation services to users who could enjoy personalized smart home control via their mobile gadgets. For instance, many of such applications would leverage the system's contextual feedback to provide users with sensing and proactive recommendations. As a result, safeguarding data communication across devices and platforms is critical to overall system security.}

    \subsubsection{\textbf{ANT-centric Security Requirements}}
    In smart home applications, it is vital to protect critical activities that communicate or store sensitive user information in their operations. As shown in Figure \ref{fig:smart home}, a typical smart home relies on a variety of smart systems and devices to provide context-aware and intelligent services to its residents. Hence, secure integration of these devices and systems is an important requirement for protecting critical data activities in smart home applications.
    From a network-centric perspective, system and device authentication, identity management, and access control are critical to prevent unauthorized access to smart home devices and to protect sensitive user information. These security requirements are crucial for safety, especially in smart home applications where unauthorized users can do harm to the owners by engaging in undesirable activities.
    From a things-centric perspective, we consider security issues that have a direct impact on smart home devices. Among all other things-centric requirements, secure updates are especially crucial for smart home applications because on-site servicing of every smart home in town will be highly inefficient and inconvenient for the inhabitants.
    
    Considering the characteristics of smart home applications, these security requirements are among the most important criteria that these applications should meet. In this part of the section, we will explain each of the requirements in detail and present some of the related proposals that can be considered for meeting these requirements.

    \paragraph{\textbf{(Activity-centric) Secure Integration}}
        A typical smart home relies on a variety of smart systems and devices to provide context-aware and intelligent services to its residents. For example, smart homes will rely on smart utility system components to offer continuous feedback on utility service consumption and system status. Such data is transformed into insights on energy trading, user consumption analytics, and even reports on leak detection. Integration of data collected from various sensors, applications, and processes is challenging, especially when smart homes have to deal with heterogeneous third-party smart household devices. 
        It is challenging to combine different technologies for efficient data collection. The context and environment in which the data value is acquired varies considerably. This adds to the challenges of delivering consistent data interpretation, and extra contextual data is required to create relevant insights. More importantly, we must address the potential security threats that such integration may bring. 
            
        Unfortunately, no examples of smart automation systems achieving system-wide security have been provided due to the high complexity involved. Nonetheless, Saleem et al. \cite{smartgridhome} have provided a good starting point where they consider the integration of the smart grid with home automation systems via a layered network structure. In this structure, a Home Area Network (HAN) is deployed in homes, connects electrical appliances with the smart meter, and is responsible for managing the power demand of residents. On top of the HAN layer, the Field Area Network (FAN) collects service and metering information from multiple HANs and transmits data to a data collector. The data collector will then exchange information with the Wide Area Network (WAN) for system-wide communication to support optimization schemes and analysis. Different application domains of these network layers have also been presented here. In addition, applications of HAN include remote home monitoring and control systems, advanced metering infrastructure (AMI) providing a two-way flow of power consumption data, integration of distributed energy resources (DERs) for better predictability and reliability of non-renewable power supply, and power demand management for consumers to reduce their utility bills. 
    
    \paragraph{\textbf{(Network-centric) Authentication}}
    
        The majority of the literature on smart home system security is centered on user authentication and identification, to prevent unauthorized access to smart IoT devices and to protect sensitive user information. Various remote user authentication scheme has been reviewed in \cite{securekeysmarthome20}, and most of the proposed scheme either fails against stolen smart card attacks or privileged insider attacks or cannot support mutual authentication. Hence, the authors proposed a secure remote user authentication scheme using symmetric encryption/decryption, XOR operations, and one-way hash functions. Although the system has been confirmed to be resistant to attacks such as device/smartphone capture attacks and user/device/gateway impersonation attacks, the use of a biometric key and a 1024-bit secret key puts severe memory constrain on the smart IoT devices. In addition, all smart IoT devices, users, and gateway nodes must complete secure offline registration with a trusted registration authority (RA), this hinders the scheme’s usability. Nonetheless, authentication in smart homes would require careful consideration of what to protect and how robust the security system should be. Following these decisions, system designers may acquire insights about feasible authentication schemes from other smart city domains.
    
    \paragraph{\textbf{(Network-centric) Identity Management and Access Control}}
        Identity management and access control refer to the technology that makes a set of rules to define who meets the requirements for accessing, using, or modifying certain resources or data in specific devices of a system. This security requirement is safety-critical in smart home applications where unauthorized users can bring damage to the owners by performing undesirable activities. For example, to avoid accidents, a visiting acquaintance should not be given access to the coffer, and children should not be given access to adult medicine.
        
        A low-cost and user-friendly smart home environment with access control is proposed in \cite{s18061886}, where they employ off-the-shelf sensors and technologies for building their smart environment. Similarly, the work in \cite{9123025} presents a practical solution for context sensitive access control in smart home systems. Their approach relies on constant monitoring of network data and detecting behavioral anomalies, and they build access control and use cases based on the knowledge graph they build. Despite the simplicity of these approaches, the presence of access control rules in both proposals will be a significant burden when a more complex smart home environment is presented.
        
    \paragraph{\textbf{(Things-centric) Secure Updates}}
    
        Remote updates are necessary for smart home applications because on-site servicing of every smart home in town will be highly inefficient and inconvenient for the inhabitants. Furthermore, in order to protect our users' safety and privacy, we must address the security problems raised by remote updates due to their wireless nature. 
        
        A combination of different security approaches can be adopted to satisfy this requirements. Besides those secure updates proposals mentioned in smart transportation, another recent direction for secure updates is the application of blockchain technology. Since the update package must maintain its integrity without compromising the code's privacy, blockchain may be a viable option for implementing secure updates. Due to the unmodifiability and transparency of blockchain, the concern for unauthorized modification will no longer exist. Any node in the distributed network can verify the validity of the update package it receives. Due to the distributed nature of blockchain, no central agency will be needed to facilitate mass updates, and it would be possible to rely on peer nodes in relaying update packages. An example of blockchain-based secure update for IoT can be found in \cite{Lee2017}.
        
        In the next section, we will investigate the challenges and gaps that current security solutions have to overcome to meet the set of security requirements from an ANT-centric perspective.

\section{Security Challenges for Smart City Applications from ANT-centric View \label{sec: challenges}}
    The interconnectedness between smart devices has a significant impact on how we live in cities, enabling a smarter, cheaper, and more convenient alternative lifestyle. These IoT-enabled smart devices could perform tasks ranging from simple lighting control to complex traffic management in metropolitan cities. Due to the vast capabilities of these smart systems and the magnitude of their impact on our lives, they require an unprecedented level of trust from their users. This trust stems from the technology's reliability and security and the system's commitment to protecting users' private assets, such as private data and accounts. McKnight et al. mentioned in one of their studies that ``one is more likely to explore and use more features of a technology if one trusts it" \cite{trustinspecifictech}.

    \textcolor{black}{With the unique set of security requirements for building smart city from the ANT-centric view, many current security proposals in section \ref{sec: context} may fail to meet them due to emerging challenges like the assumption of trusted cloud server.} While numerous proposals have been made for secure smart city applications, most schemes that use contemporary innovations, such as blockchain, cloud and edge servers, need to focus more on one or more security requirements. Furthermore, with growing connectivity between heterogeneous devices and systems in smart city applications, developers have to address a new set of security challenges in addition to the traditional security challenges.
    \textcolor{black}{In this section, we present the set of challenges that current security solutions for smart city applications will encounter in meeting this set of requirements and provide a new research perspective towards the evaluation of security frameworks and design. These could also serve as future research directions where research can be aimed at fulfilling these challenges in existing frameworks or giving insights into future research directions.
    }
    
    \textcolor{black}{To better illustrate these security challenges, we categorize them according to the ANT architecture.} In this view, the activity-centric perspective facilitates understanding of system components' context and identifies security challenges relating to the sensitive operations performed in the system. In comparison, the network-centric perspective focused on security challenges for wireless communication between network components and the flow of sensitive data. The things-centric perspective aids in understanding the relationship between the security challenges of heterogeneous data acquisition via heterogeneous IoT devices.

    \subsection{Activity-Centric Security Challenges}
    The activity-centric perspective facilitates understanding of the context of system components and identifies security challenges relating to the sensitive operations performed in the system. We have identified a few security challenges related to the usage of edge/cloud systems and challenges originating from big data and the use of big data in our operations, which are more relevant to IoT systems in smart cities in this subsection.
    
    \subsubsection{\textbf{Assumption of Trusted Cloud/Edge}}
        The adoption of cloud infrastructure to improve the efficiency and performance of our systems is a key feature of the 5G network \cite{5g6g, 5g}. In many smart city applications, the edge cloud is essential for bridging the central cloud and the local user to facilitate efficient data exchanges. Similarly, the adoption of edge servers for storage or computation offloading has become popular due to its relatively low latency and high computation capabilities.
        
        However, a significant challenge for such applications is the assumption that the cloud/edge can always be trusted. Coppolino et al. \cite{cloudsecurity17} have identified the main security challenges in the cloud, including shared technologies vulnerability, data breach, account or service traffic hijacking, DDoS, and malicious insiders. To guard against data theft, system designers have to consider major players, such as external attackers, malicious internal users, and even hostile employees from cloud providers. Furthermore, solutions employing hybrid cloud and edge cloud also have to consider intercloud interactions' security to ensure consistent data protection throughout the backend process.

    \subsubsection{\textbf{Accommodating Enormous Data Volume and Dynamic Traffic}}
        With the deployment of smart IoT sensors in cities, large amounts of data are continuously generated, and this data forms the backbone of urban intelligence. This data load poses a significant challenge to the current technological infrastructure in processing, storing, and providing real-time insights. As a result, resource allocation and system performance optimization are critical for systems to meet their expectations in the face of large data volume. In this context, any system attempting to address smart city issues needs to be prepared to handle a non-periodic, sudden burst of data gracefully. Depending on the application, this could entail efforts to maintain timely service delivery in a time-sensitive application or handle an unanticipated burst in data volume without compromising service quality. Similarly, the security mechanism must accommodate an instantaneous surge in traffic without compromising security or performance, guaranteeing that the quality of security services promised to the user is delivered.

    \subsubsection{\textbf{New Types of Interactions}}
        Different smart city systems could potentially interact due to the growing connectivity between them, leading to the frequent occurrence of cross-system or cross-platform interactions. Thus, a proliferation of new interaction types and styles can be expected. We will need to specify rules for handling such non-traditional interactions and incorporate resilience into our systems to protect against them. In addition, system designers have to consider the system's openness and account for the security risk for possible interactions with external platforms or users.
        
        Consider the case of D2D wireless communication resource allocation in IoT applications, where there is a high density of smart devices per unit area that constantly demands communication resources. According to Hussain et al., \cite{iOtresouceallocation20}, this leads to congestion, network overload, and degradation of the Signal Noise Ratio in the wireless communication channel. While the traditional centralized approach deals with large but infrequent data transmission from traditional communication devices, IoT devices require a new approach for small and frequent data exchanges. Additionally, while frequent packet transmission might be a sign of flooding in many traditional systems, many IoT device interactions are characterized by small and frequent data transmission; therefore, accessing such metrics the same way as conventional systems would be inappropriate.
    
        Similarly, the temporal, once-only interaction with unknown smart devices can disturb system security in smart city scenarios. In the case of IoV, a user might travel to a neighboring country and share data with other IoV or smart transportation infrastructures there. The two transportation systems may not use a single platform and protocol due to different government regulations. However, to fully utilize the capabilities of IoV, the car must interface with other devices to gather traffic intelligence. While most of these interactions will be once-off, it is necessary to provide security protection to ensure the security of the device itself, i.e., not to be hijacked by others or attacked.

    \subsubsection{\textbf{Lack of Unified Representation}}
        A unified representation of heterogeneous data types can support effective insight generation across devices and systems. It also allows systems to access the context and the security risk of the data exchange. Thereby, a more appropriate security setting can be established, avoiding unwanted interactions with sensitive data and preventing data loss. The problem of defining a unified representation of heterogeneous data sources for cooperative tasks has been intensified by the introduction of big data. Prior to the emergence of big data, researchers have been focusing on knowledge discovery for structured data, attempting to unify representation among structured databases. For instance, the authors in \cite{globalview01} have studied the problem of data integration for the structured data source. The types of semantic heterogeneity for structure data studied in the paper include name and structural heterogeneity, which can be further subdivided into type conflict, key conflicts, and domain conflicts. 
        
        Besides the heterogeneity mentioned above in structured data, the author in \cite{bigdataheterogeneity17} has studied the heterogeneity in big data, which is likely to be a major characteristic for smart city applications. The heterogeneity in big data includes syntactic heterogeneity, conceptual heterogeneity, terminological heterogeneity, and pragmatic heterogeneity. Although advanced processing and analysis techniques can achieve data integration between structured and machine data to a reasonable extent, providing a high-level data fusion between unstructured analog sensor data will remain a challenging task. However, due to the complexity of such tasks, it is tough to achieve a context-independent approach to present a unified representation of the heterogeneous data. As a result, appropriately handling heterogeneous data types may be a significant security challenge for smart city system designers.

    \subsubsection{\textbf{Data Sensitivity}}
        Different data types have different sensitivity levels, and even the input data in an application can have different levels of sensitivity. Pulse data, mobility tracker data, and a live video of the patient may all be fed into an in-home health monitoring system simultaneously. From a privacy-preserving perspective, the patient's live video data deserves the most stringent security mechanism it could provide. From a healthcare product perspective, pulse data should remain top-secret to meet customers' expectations on the security of their health data. After defining the sensitivity of input data, how different data types should be protected comes the question. While we are familiar with how structured data could be stored and protected, the balance between protection for unstructured data and the ease in extracting knowledge from it for integrated information queries remains difficult. Thus, designing system security architecture with data of different sensitivity will be a challenging task.

    \subsection{Network-Centric Security Challenges}
        Secure communication is the foundation of a secure system. It safeguards the confidentiality, integrity, and availability of information exchanged in the network. Common attacks on wireless communications channels include blackhole, wormhole, and pharming attacks that disrupt the delivery of data; Jamming channel, DDoS, and Syn Flooding attacks that disable the availability of services; eavesdropping and Traffic Analysis that leaks confidential data to unauthorized third parties. Thus, prevention and detection of attacks on wireless communication channels are critical for all systems relying on wireless networks to perform data transmission to provide reliable and timely service delivery. In addition, the survey in \cite{wirelesscom18, landaluce2020review} has presented a few communication security challenges prevalent in IoT systems that deserve more attention. For example, the size of cryptographic keys and certificates for trust management must be kept short due to resource constraints in IoT devices, and the performance overhead for secure storage or code attestation must be within reasonable parameters to extend the battery life of devices that are not powered directly.
        Furthermore, while the standardization of network protocols is extremely challenging due to different performance requirements between different system components, secure and efficient routing is vital to the overall security of the system. If more than one routing protocol is adopted throughout the system, the compatibility between protocols should also be investigated.

        \subsubsection{\textbf{Wireless Communication Protocols}}
            A set of devices communicating with neighbors within its radio range can form a network, which can be arranged in different topologies. Simple network topology can achieve low latency in relaying messages, while complex network topology can offer better routing decisions and flexibility \cite{wirelesscom18}. There are four main network topologies that are commonly implemented: 
            
  
            
            \begin{enumerate}
            \item \textbf{Star Topology:} In the star topology, all nodes are in the radio range of the central node and communicate directly with the central node. All network traffic in this network will pass through the central node. This is the most simple topology and works best for typical master-and-slave relationships between devices.
            \item \textbf{Tree Topology:} In the tree topology, each node is within the radio range of its parent branch and children branch. In this topology, network communication can be multi-hoped, with nodes relying on their neighbors to relay messages to nodes outside of their radio range.
            \item \textbf{Mesh Topology:} In the mesh topology, each node can establish a wireless communication link with one or more neighbors, allowing for more traffic routing options in the network communication. However, as the number of possible routing paths grows, the complexity of the routing decision increases as well.
            \item \textbf{Cellular Topology:} In the cellular topology, each node communicates directly with the cluster head of their cell. These cluster heads are placed so that they can offer sufficient coverage for all nodes while avoiding radio range overlaps with other cluster heads. For inter-cluster communication, these cluster heads communicate with other cluster heads directly.
            \end{enumerate}

            Communication protocols are rules that define how data is exchanged within the network. Many wireless IoT communication protocols have been developed for various uses due to the various device capabilities and requirements. Table \ref{tab: protocol} provides a description of the security features of the most commonly used wireless communication protocols for smart city applications, as well as their corresponding security challenges. While there is no standardization for the protocols, data formats, and systems adopted by vendors, it is common to see smart devices of different types and brands being deployed in the same system. Hence, interoperability between devices becomes a major challenge in providing secure communication between heterogeneous devices.

        \subsubsection{\textbf{Security Protocols}}
        Various standard security protocols and solutions have been developed to ensure communication security, such as DTLS, CETIC-6LBR(6LoWPAN border router), IPsec, IKEv2, and Host Identity Protocol (HIP). The authors in \cite{wirelesscom18} present a summary of wireless communication principles about the connectivity requirements of IoT. Communication standards and the gap between their security features and actual IoT system vulnerabilities are also highlighted. 
        
        For example, traditional security protocols are built on OSI stacks, which may or may not be fully implemented in many IoT systems. Traditional security mechanisms are often employed in the Medium Access Control (MAC) or higher layers, assuming that the underlying physical link and data link layers will provide a reliable system link. However, the physical layer security follows the information-theoretic principle. The strength of the security mechanism depends on the assumption of varied channel conditions between the user and attacker. As a result, security guarantees are rarely 100 percent. Furthermore, the system architecture of many IoT systems differs significantly from the traditional system architecture, where they introduce their layers, such as perception and actuation layers. These new layers often require new adaptions of current security protocols to ensure system security. Hence, it is wise to investigate the security requirements for every layer of the system and their context to evaluate the choice of security protocols used for system security.
            
        \subsubsection{\textbf{Distributed Communication}}
        The majority of current wireless IoT communication solutions use a centralized approach, in which a central authority or server manages data sharing between devices. As the number of devices in the network grows, communication to and from the center server will quickly become a bottleneck, resulting in substantial network latency. While others argue that edge servers should be used to offload centralized computing activities, a decentralization approach can best deal with the issue of scalability. One of the ways to establish a decentralized IoT network is via P2P networking technologies. Like any other widely adopted technology, when P2P technology is commonly deployed, we expect many adversarial attacks. 
        
        In \cite{p2psecuritysurvey2003}, the authors highlight that besides adversarial attacks, there are other challenges in the architecture of a P2P network. One such issue is the selfish behavior of users. While many P2P applications are designed to serve a large number of users over a network, a ``hotspot" issue may nevertheless arise, in which some nodes with popular content must service more peers than others. Some selfish users might want to reserve more resources for themselves, such as disk space, bandwidth, and power, by altering their system configurations to act in their interest. Although these users are not explicit attackers, their action degrades the overall system performance if no checks and balance are in place. Another significant challenge in a peer-to-peer network is the concept of trust. Since there is no root of trust, the system must devise a way to assess the trustworthiness of external data transfers and respond appropriately to defend itself. Thus, providing a secure approach for distribution communication is a challenging task.

    \subsection{Things-Centric Security Challenges}
        Due to the growing connectivity between systems and applications' dependence on big data, many smart city applications will be directly or indirectly dependent on a large scale of IoT devices for data collection. Unfortunately, while these IoT devices are important data sources that support system operations, many of them have limited resources and capacity to implement comprehensive security measures, leaving them vulnerable to attack. More malicious attempts by attackers probing for the weakest link in the system are likely to occur due to the large-scale physical deployment, which increases the attack surface for attackers to interfere with the systems. 
        
        At the same time, due to the massive deployment scale, a successful intrusion into these end devices might cause a tidal effect, compromising the security of the system and even paralyzing it. Consequentially, the effect of intrusion will be very significant and impacts many devices. In addition, dealing with the aftermath of an attack can be equally concerning. Traditionally, technicians could upgrade the sensor network on-site, either to install patches for updates or to address a loophole in these devices. Such a management approach is no longer feasible, especially for mass upgrades, due to a large number of devices deployed. Furthermore, many of these devices may be physically inaccessible to technicians or almost impossible to access. Hence, we shall direct our interest towards remote updating mechanisms, performing updates, and integrity attestation over the air. However, due to the wireless nature of such mechanisms, security concerns have become growingly prominent. Thus, smart city systems that have a large-scale physical deployment of devices in open environments will therefore contribute to a proliferation of security challenges, such as ensuring secure mass updates and performing large-scale system recovery.

\section{Future IoT and Future IoT Security \label{sec: future}}
    \textcolor{black}{As our vision of a smart city becomes a reality, it is essential that our security techniques grow in tandem with the changing landscape.} Future IoT in smart cities is expected to be fully intelligent, with seamless cross-functional and cross-platform interactions. An important characteristic of the data communication in future smart city application is wireless. While a small portion of current applications can maintain their usability using wired data communication, that is not feasible for city wide deployment of intelligent systems. Furthermore, wireless communication protocols can promote effective communication between various systems, where each wireless component of the systems might possibly operate as a point of interaction between systems.
    
    As IoT progresses from ``perceptive" to ``cognitive," we expect these devices to learn from our actions and help pave the way for deeper integration of people, things, and infrastructure. Cognitive IoT devices can sense their surrounding environment, gathering information about other participants nearby. The intelligence agent is fed with relevant data from users and the environment and a corresponding set of execution options and system/environmental feedback from the execution. The more information the cognitive IoT device can collect, the better it could formulate its decision path. By iteratively updating the best strategy under user-defined policies to achieve the best reward, the agent gains the ability to make intelligent decisions. Since the core model's cognitive ability is based on the collection of behavioral insights from our actions and feedback, which serve as the model's knowledge base, the quality of data can heavily impact the model's performance. 
    
    However, since the cognitive IoT device can collect all forms of data and infer high-level interpretation, it is essential to construct a new set of regulations to limit the types of data collected. While devices can record high-precision photos and contribute to the formation of high-resolution road maps, they can also catch imagery of other persons and objects. With strong cognitive abilities and high computation capabilities, the device could run facial recognition models and search for the internet presence of the passerby. As neural networks become more powerful, they may one day reveal a great deal of personal information about strangers from the data collected.
    
    As IoT applications become more intelligent, new security vulnerabilities emerge. To begin with, such a model's training necessitates a large database, and the model could allow for online training, where new samples generated in real-time can be processed and fed into the model for training. The high volume and ever-expanding dataset put enormous strain on the system's ability to store and analyze data securely. Sharing and analyzing of cyber threat intelligence files between different security organizations in smart cities to maintain awareness of emerging vulnerabilities is also necessary \cite{yang2019automated}. In addition, since more cross-functional interactions are anticipated, we can expect more frequent and massive data flows between systems, exposing more attack surfaces to external attackers. Future IoT security would require extensive protection towards sensitive data flow among system components and across systems. Furthermore, as the number of IoT devices grows exponentially, more systems will be encountering large traffic flow from these devices due to the increasing connectivity between systems. The large growth in IoT devices further exacerbates the demand for wireless bandwidth. To achieve better efficiency with large data traffic, a popular approach is the adoption of cognitive radio, which provides appealing strategies such as spectrum sensing, self-learning, circumstantial perceiving, and spectrum access. However, they can also add to security and privacy risks \cite{Advancedandemerging20}. 
    
    Cognitive IoT devices can perform spectrum sensing to detect ``holes" in the spectrum and monitor the behavior of network users. This can lead to two significant issues. Firstly, since cognitive IoT devices can monitor the behavior of network users, the usages patterns of these network users can be tracked, thus resulting in user privacy concerns. Without a proper framework for secure and efficient spectrum trading, spectrum sensing by unauthorized devices might be considered a pre-event sign of intrusion in the wireless network.
    
    On the other hand, the machine learning \cite{ml} approach could also be applied in IoT security, providing systems with automated/semi-automated security defense. In \cite{8454402}, the author discusses how machine learning may be utilized for learning-based authentication, access control, IoT offloading for addressing attacks on the physical and MAC layers, and IoT malware detection. The proposed methods utilize machine learning techniques as the decision-making backbone, with large models and networks serving as the decision-making backbone. Although present IoT systems cannot handle such computing overhead and cannot provide sufficient state observation to train such models, future IoT security can aim to achieve a good balance of intelligence and performance with the help of machine learning. Besides security and privacy challenges, the standardization of cognitive IoT networks remains a challenging task. While much research has been done in this area, most of the solutions address part of the issue. More research is needed to integrate these proposals and provide proper standardization.

\section{Conclusion \label{sec: conclusion}}

In this paper, we have introduced a new perspective on security issues in smart cities from a data-centric viewpoint by referring to ANT-centric architecture. We studied the concept of a smart city holistically, starting with end applications to abstract shared characteristics and highlighted some privacy and security challenges these smart city applications will face. Following that, we evaluated the specific context of each smart city domain and the context-specific security requirements. By doing so, we want to provide directions for smart city system designers on when, where, and how they should implement security strategies and the possible security challenges they need to consider. Furthermore, we investigated the challenges and gaps that current cryptographic solutions have to overcome. Lastly, we give our view of the future IoT and IoT security.
    
Overall, while emerging technologies have enormous potential for addressing difficulties in IoT-enabled smart cities, security must be considered throughout the data life cycle and any possible interactions with the outside world. Only then will the system withstand attacks on every component of the large network. Aside from the usual security considerations, proposals for IoT-enabled smart city applications must carefully evaluate the application's context and social expectations to ensure its feasibility. Other security issues such as secure updates for massive IoT networks, interoperability between different protocols, and maintenance of Quality of Security Services deserve more research attention. As our IoT systems gradually evolve from ``perceptive" towards ``cognitive", autonomous or semi-autonomous complex systems capable of cognitive decision making will become commonplace. How best to implement security in such a sophisticated and intelligent system will be of great value towards the development of our smart city landscape.

\bibliographystyle{bibstyle}
\bibliography{references}

\vskip -1\baselineskip plus -1 fil
\begin{IEEEbiography}[{\includegraphics[width=1in,height=1.25in,clip,keepaspectratio]{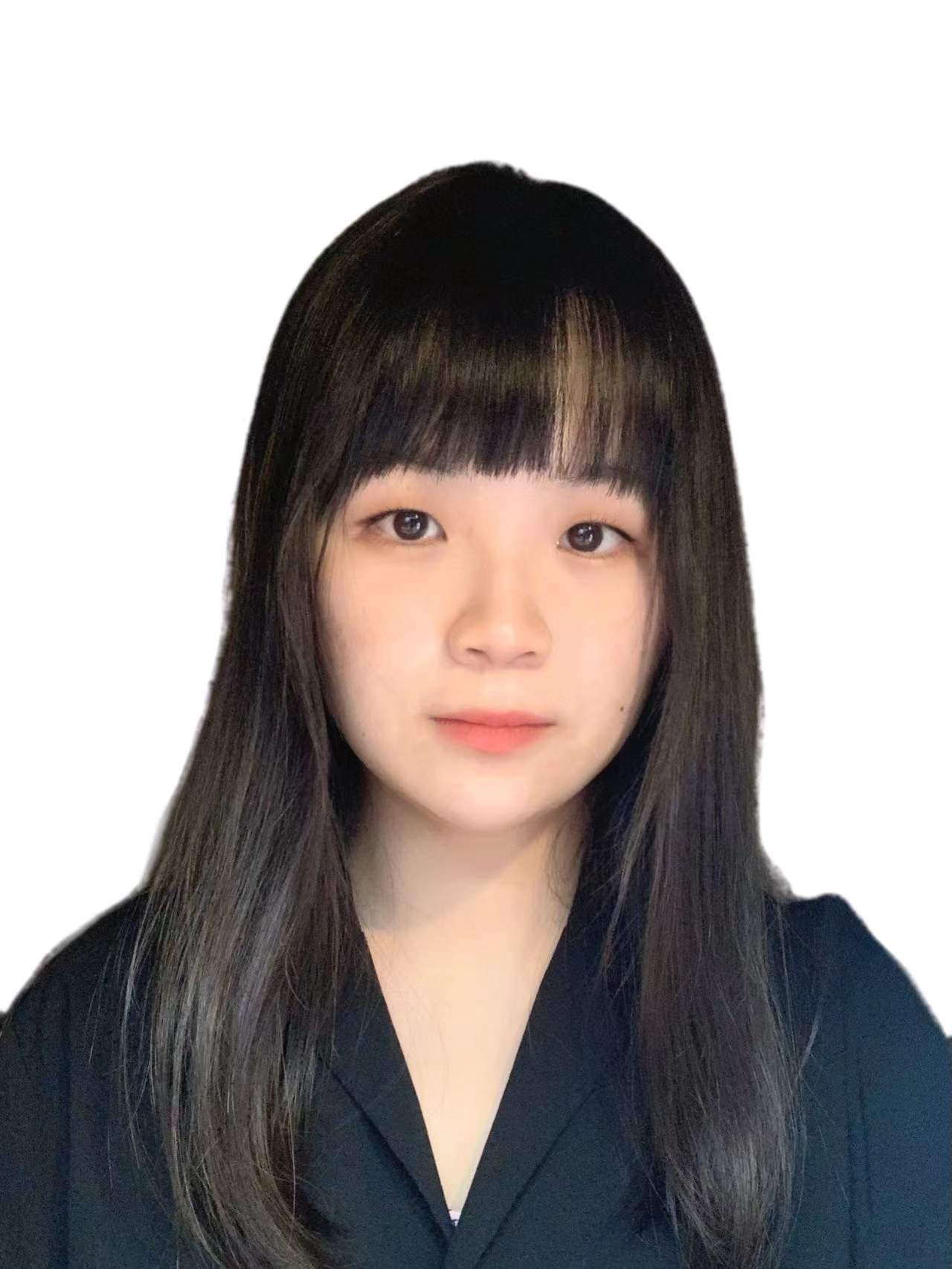}}]{Jiani Fan}received the Bachelor’s degree in information systems from Singapore Management University in 2020. She is currently pursuing the Ph.D. degree in computer science at Nanyang Technological University. Her research interests include IoT Security, Cybersecurity, and Internet of Vehicles.
\end{IEEEbiography}

\vskip 0pt plus -1 fil
\begin{IEEEbiography}[{\includegraphics[width=1in,height=1.25in,clip,keepaspectratio]{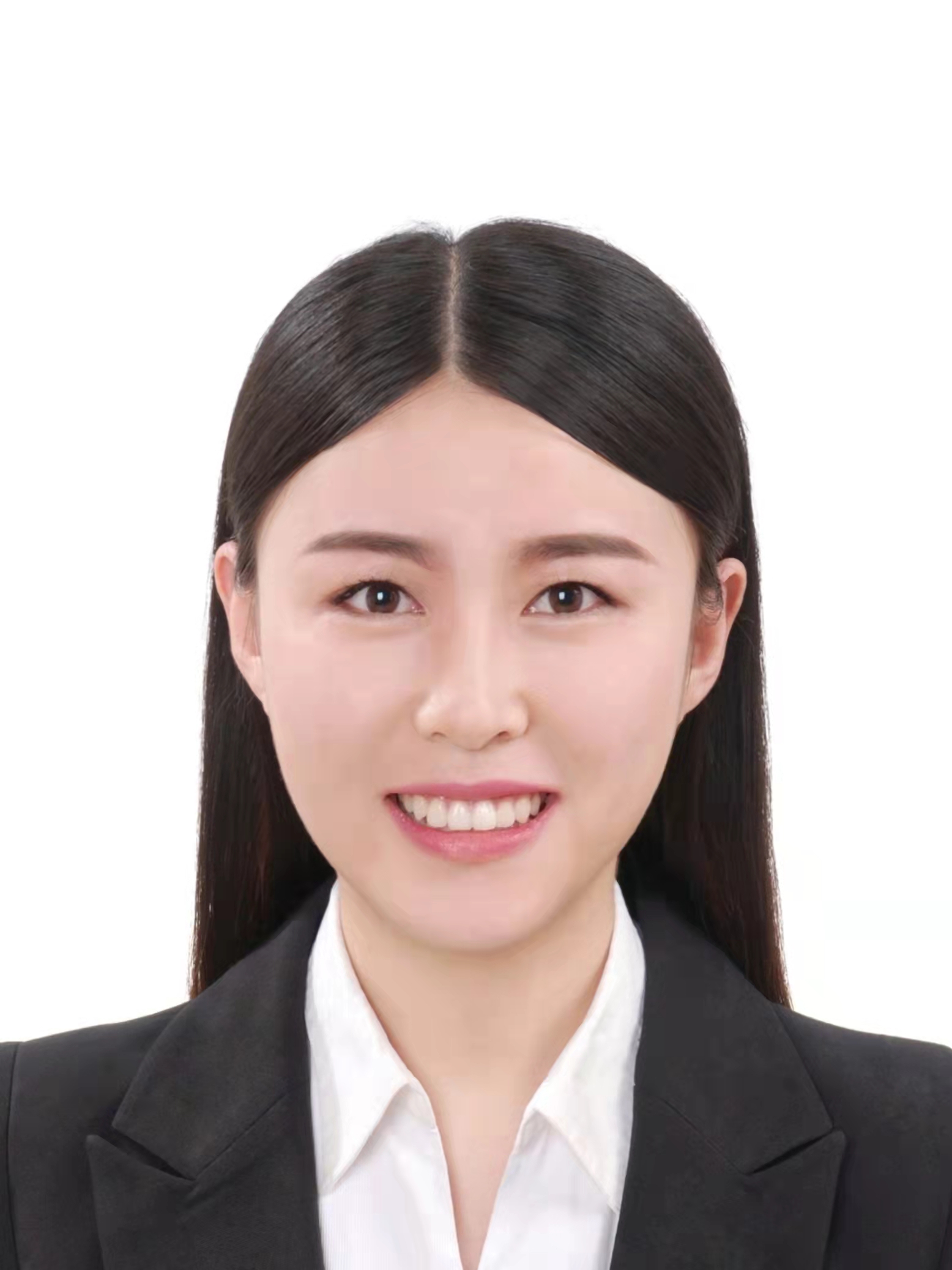}}]{Wenzhuo Yang} received the Bachelor’s degree in Measuring and Controlling Technologies and Instruments from Beijing University of Posts and Telecommunications, Beijing, China, in 2016. She is currently pursuing the Ph.D. degree with the School of Computer Science and Engineering, Nanyang Technological University, Singapore. Her research interests include Privacy-Preserving Machine Learning, IoT Security, Intrusion Detection, and Cyber Threat Intelligence Analysis.
\end{IEEEbiography}

\vskip 0pt plus -1 fil
\begin{IEEEbiography}[{\includegraphics[width=1in,height=1.25in,clip,keepaspectratio]{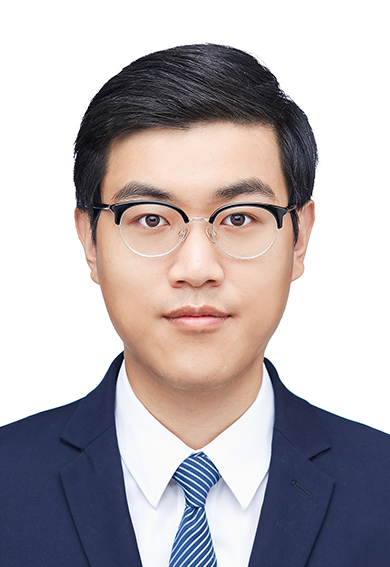}}]{Ziyao Liu}
received his B.E. degree from the school of Electronics Information Engineering, Zhengzhou University, Zhengzhou, China, in 2015, and the M.S. degree from Beijing Institute of Technology, Beijing, China, in 2018. He is currently working towards a Ph.D. degree in the School of Computer Science and Engineering, Nanyang Technological University, Singapore. His research interests include privacy-preserving machine learning, multi-party computation, and applied cryptography.
\end{IEEEbiography}
\vskip 0pt plus -1 fil

\begin{IEEEbiography}[{\includegraphics[width=1in,height=1.25in,clip,keepaspectratio]{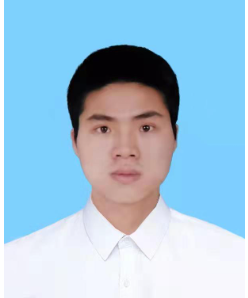}}]{Jiawen Kang} received the Ph.D. degree from the Guangdong University of Technology, China in 2018. He was a postdoc at Nanyang Technological University, Singapore from 2018 to 2021. He currently is a professor at Guangdong University of Technology, China. His research interests mainly focus on blockchain, security, and privacy protection in wireless communications and networking.
\end{IEEEbiography}

\vskip 0pt plus -1 fil
\begin{IEEEbiography}[{\includegraphics[width=1in,height=1.25in,clip,keepaspectratio]{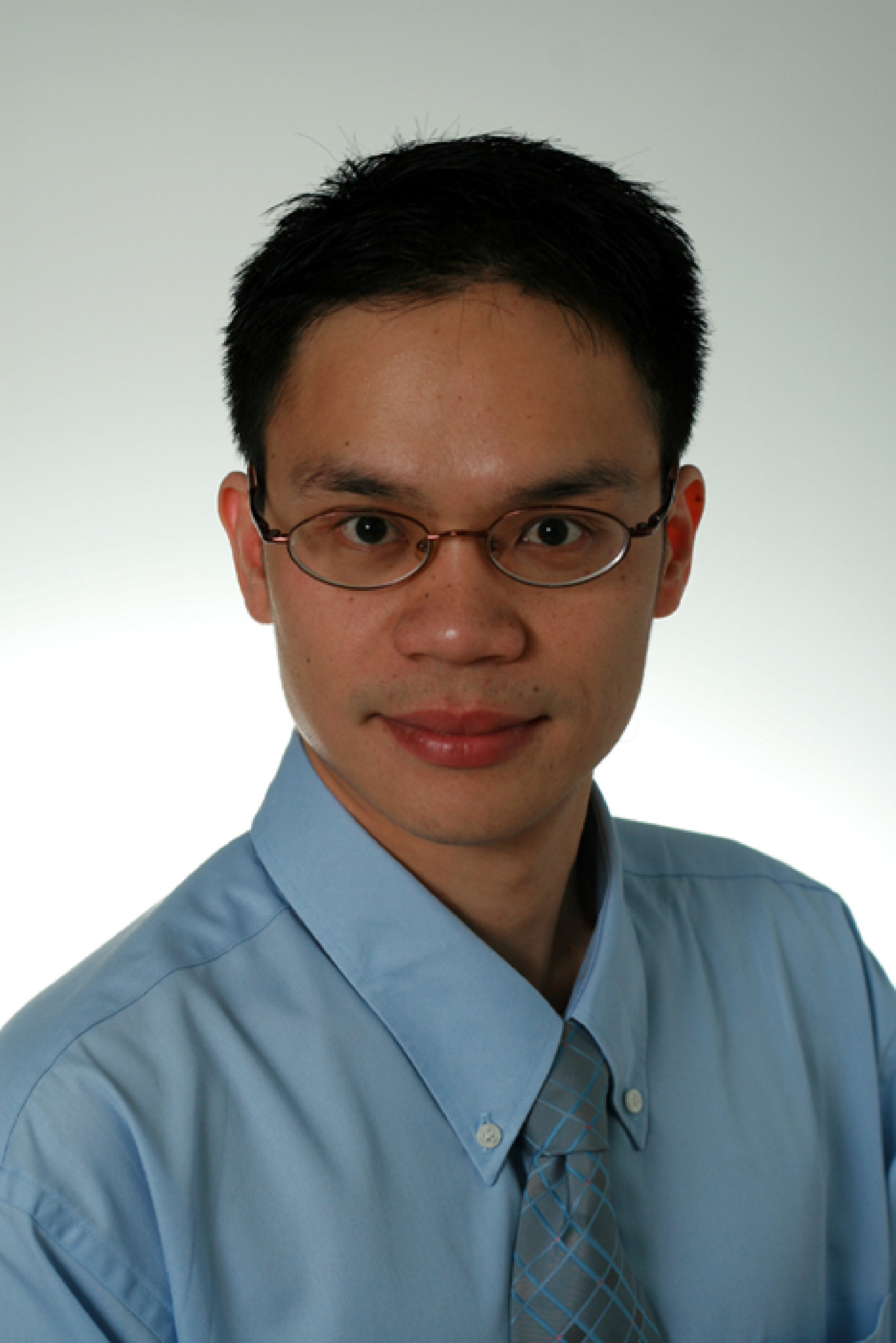}}]{Dusit Niyato} (Fellow, IEEE) is currently a professor in the School of Computer Science and Engineering, at Nanyang Technological University, Singapore. He received B.Eng. from King Mongkut's Institute of Technology Ladkrabang (KMITL), Thailand in 1999 and the Ph.D. in Electrical and Computer Engineering from the University of Manitoba, Canada in 2008. His research interests are in the areas of the Internet of Things (IoT), machine learning, and incentive mechanism design.
\end{IEEEbiography}

\vskip 0pt plus -1 fil
\begin{IEEEbiography}[{\includegraphics[width=1in,height=1.25in,clip,keepaspectratio]{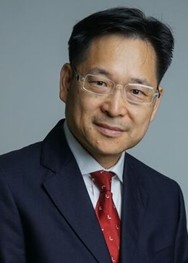}}]{Kwok-Yan Lam}
(Senior Member, IEEE) received his B.Sc. degree (1st Class Hons.) from University of London, in 1987, and Ph.D. degree from University of Cambridge, in 1990. He is the Associate Vice President (Strategy and Partnerships) and Professor in the School of Computer Science and Engineering at the Nanyang Technological University, Singapore. He is currently also the Director of the Strategic Centre for Research in Privacy-Preserving Technologies and Systems (SCRiPTS). From August 2020, he is on part-time secondment to the INTERPOL as a Consultant at Cyber and New Technology Innovation. Prior to joining NTU, he has been a Professor of the Tsinghua University, PR China (2002–2010) and a faculty member of the National University of Singapore and the University of London since 1990. He was a Visiting Scientist at the Isaac Newton Institute, Cambridge University, and a Visiting Professor at the European Institute for Systems Security. In 1998, he received the Singapore Foundation Award from the Japanese Chamber of Commerce and Industry in recognition of his research and development achievement in information security in Singapore. His research interests include Distributed Systems, Intelligent Systems, IoT Security, Distributed Protocols for Blockchain, Homeland Security and Cybersecurity.
\end{IEEEbiography}

\vskip 0pt plus -1 fil
\begin{IEEEbiography}[{\includegraphics[width=1in,height=1.25in,clip,keepaspectratio]{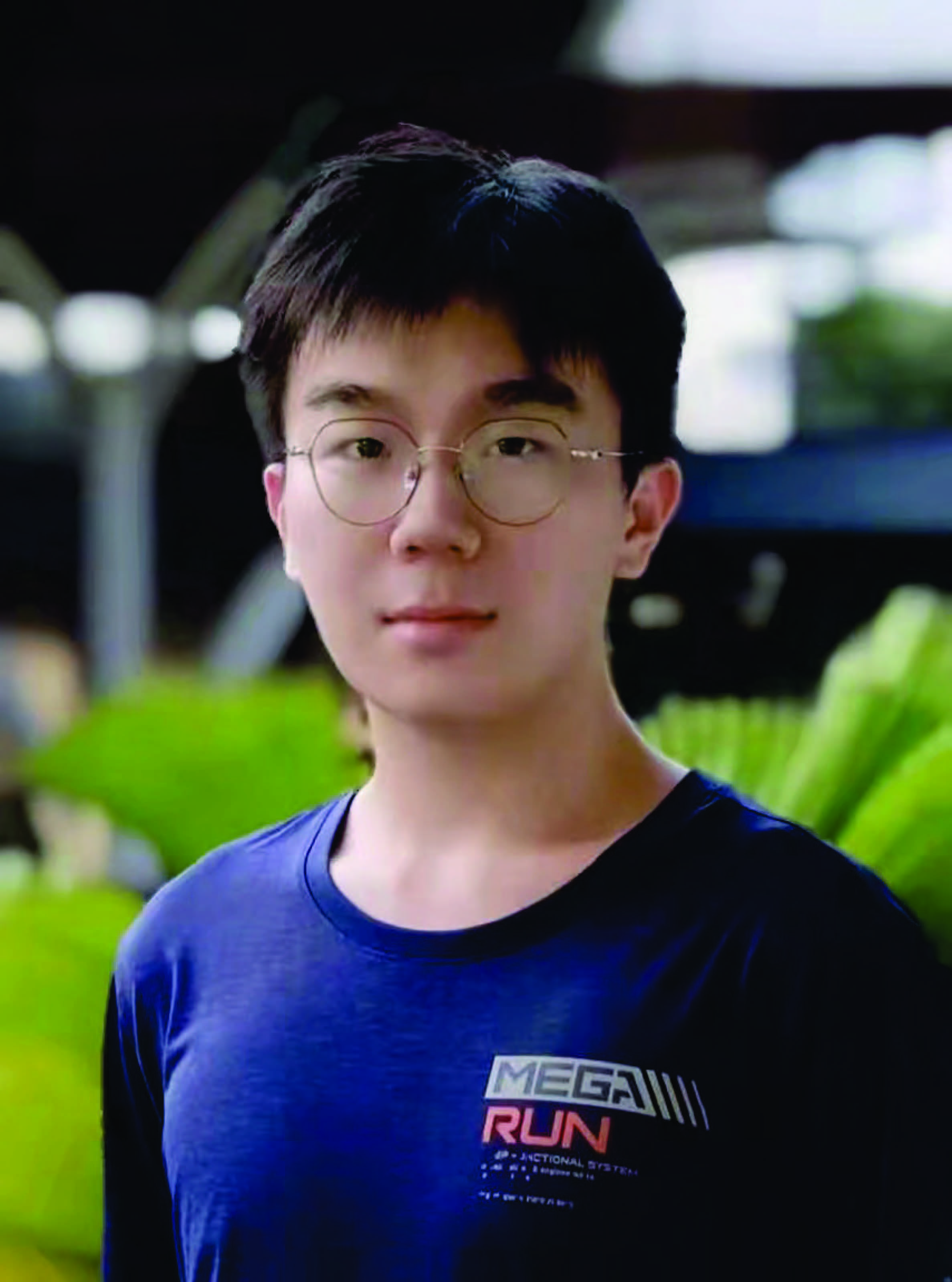}}]{Hongyang Du}
received the B.Sc. degree from Beijing Jiaotong University, Beijing, China, in 2021. He is currently pursuing the Ph.D. degree with the School of Computer Science and Engineering, the Energy Research Institute @ NTU, Interdisciplinary Graduate Program, Nanyang Technological University, Singapore. He was recognized as an exemplary reviewer of the \textsc{IEEE Transactions on Communications} in 2021. His research interests include reconfigurable intelligent surface and communication theory.
\end{IEEEbiography}

\end{document}